\documentclass[aps,prd,nofootinbib,onecolumn,a4paper]{revtex4-2}
\usepackage{bm,latexsym,amsmath,amssymb,amsfonts,mathrsfs,amsfonts}
\usepackage[dvipdfmx]{graphicx}
\usepackage[hang,small,bf]{caption}
\usepackage[subrefformat=parens]{subcaption}
\usepackage{ascmac}
\usepackage{physics}
\usepackage{color,float}
\usepackage{braket}
\usepackage{bbm}
\usepackage{comment}

\newcommand{\simgt}{\lower.5ex\hbox{$\; \buildrel > \over \sim \;$}}
\newcommand{\simlt}{\lower.5ex\hbox{$\; \buildrel < \over \sim \;$}}

\begin{document}

\begin{flushright}
  KEK-TH-2566
\end{flushright}

\title{\bf  Violation of the two-time Leggett-Garg inequalities  for a harmonic oscillator}

\author{Kosei Hatakeyama}
\affiliation{Department of Physics,  Kyushu University, 744 Motooka, Nishi-Ku, Fukuoka 819-0395, Japan}

\author{Daisuke Miki}
\affiliation{Department of Physics,  Kyushu University, 744 Motooka, Nishi-Ku, Fukuoka 819-0395, Japan}

\author{Masaki Tani}
\affiliation{Department of Physics,  Kyushu University, 744 Motooka, Nishi-Ku, Fukuoka 819-0395, Japan}

\author{Yuuki Yamasaki}
\affiliation{Department of Physics,  Kyushu University, 744 Motooka, Nishi-Ku, Fukuoka 819-0395, Japan}

\author{Satoshi Iso}
\affiliation{Theory Center, High Energy Accelerator Research Organization (KEK), Oho 1-1, Tsukuba, Ibaraki 305-0801, Japan}
\affiliation{
Graduate University for Advanced Studies (SOKENDAI),
Oho 1-1, Tsukuba, Ibaraki 305-0801, Japan}
\affiliation{
International Center for Quantum-field Measurement
Systems for Studies of the Universe and Particles (QUP),
KEK, Oho 1-1, Tsukuba, Ibaraki 305-0801, Japan}

\author{Apriadi Salim Adam}
\affiliation{Research Center for Quantum Physics, National Research and Innovation Agency (BRIN), South Tangerang 15314, Indonesia}

\author{Ar Rohim}
\affiliation{Departemen Fisika, FMIPA, Universitas Indonesia, Depok 16424, Indonesia}

\author{Kazuhiro Yamamoto}
\affiliation{Department of Physics,  Kyushu University, 744 Motooka, Nishi-Ku, Fukuoka 819-0395, Japan}
\affiliation{
Research Center for Advanced Particle Physics, Kyushu University, 744 Motooka, Nishi-ku, Fukuoka 819-0395, Japan}
\affiliation{
International Center for Quantum-field Measurement
Systems for Studies of the Universe and Particles (QUP),
KEK, Oho 1-1, Tsukuba, Ibaraki 305-0801, Japan}

\date{\today}

\begin{abstract}
We investigate the violation of the Leggett-Garg inequalities for a harmonic oscillator in various quantum states. We focus on the two-time quasi-probability distribution function with a dichotomic variable constructed with the position operator of a harmonic oscillator. 
First, we developed a new formula to compute the two-time quasi-probability distribution function, whose validity is demonstrated in comparison with the formula developed in the recent paper by Mawby and Halliwell~[Phys.~Rev.~A~{\bf 107}~032216 (2023)].
Second, we demonstrated the variety of the violation of the two-time Leggett-Garg inequalities assuming various quantum states of a harmonic oscillator including the squeezed coherent state and the thermal squeezed coherent state.
Third, we demonstrated that a certain type of extension of the dichotomic variable 
and the corresponding projection operator can boost violation of the Leggett-Garg inequalities for the ground state and the squeezed state. 
We also discuss when the Leggett-Garg inequalities are violated in an intuitive manner. 
\end{abstract}
\maketitle

\section{Introduction}
Testing quantum coherence in macroscopic systems is one of the  fundamental problems in modern physics to explore the boundary 
between the quantum world and the classical world.
The Leggett-Garg inequalities are the relations that must be satisfied from two intuitive principles in the macroscopic world \cite{Leggett85,Leggett02,Emary}:
macrorealism and noninvasive measurability.
Macroscopic realism means that the physical quantity is a predetermined value regardless of the measurements.
The second principle implies that we can measure this predetermined value without disturbing the system's state.
In contrast, quantum mechanics breaks these two principles because of quantum superpositions and state collapse.
Experiments to verify the violation of the Leggett-Garg inequalities have been performed on spin operators in qubit systems and superconducting circuits \cite{Knee,Ruskov,Palacios,Xu,Dressel}.
Recently, they were also applied to neutron interferometers to test how far the prediction of quantum mechanics holds against macrorealism \cite{neutron}. The coherence of the neutrino oscillation was tested using the violation of the Leggett-Garg inequalities \cite{neutrino}.
These are the frontiers of testing quantum mechanics in the macroscopic world. There are further proposals, e.g., testing the quantum nature of macroscopic systems with the Leggett-Garg inequalities including the gravity \cite{MNY}.

The authors of Ref.~\cite{Bose} demonstrated that the Leggett-Garg inequalities can be
violated in a harmonic oscillator in coherent states. 
Furthermore, Halliwell's group has investigated the violation of the 
Leggett-Garg inequalities in harmonic oscillators \cite{
Halliwell17,Halliwell19,Halliwell21,Halliwell22, Halliwell23}.
The optomechanical oscillator system is a promising method for preparing the quantum states of a massive object as tabletop experiments 
\cite{Aspelmeyer,Bowen,Yambei,Michimura}. Continuous measurement cooling technology has been developed to realize such a quantum state \cite{Matsumoto,MY,Meng}.
Feasibility tests have demonstrated that the quantum states of a macroscopic pendulum will be realized in the near future \cite{Miki3,Sugiyama,Shichijo}.
As a first step toward a test of the macrorealism with a macroscopic oscillator, we investigate the violation of the Leggett-Garg inequalities with a harmonic oscillator in various quantum states. 
The purposes of the present paper are as follows: 
The first one is to develop a new formula to compute the two-time quasi-probability distribution function, whose validity will be demonstrated in comparison with the method developed in the recent paper by Mawby and Halliwell \cite{Halliwell23}.
The second one is to demonstrate the violations of the two-time Leggett-Garg inequalities explicitly assuming the various quantum states of a harmonic oscillator including the squeezed coherent state and the thermal squeezed coherent state. 
The third one is to demonstrate a simple extension of the dichotomic variable and the projection operator, 
which shows the violation of the Leggett-Garg inequalities for the ground state and the squeezed state, where the violation can be boosted by the simple choice of the projection operator. 
Based on these investigations, we discuss when the violation of the Leggett-Garg inequalities occurs intuitively.

The present paper is organized as follows: 
In Sec.~2, we briefly review the basic formulas of the two-time quasi-probability distribution function for the two-time Leggett-Garg inequalities. We develop a new formula to compute the quasi-probability distribution function useful for quantum continuous variables by using the integral representation of the Heaviside function. 
In Sec.~3, We also find the expression for the two-time quasi-probability distribution function by extending the formulas developed in Ref.~\cite{Halliwell23}. We compare the results with these two different formulas. 
We also investigate the behaviors of the quasi-probability distribution function of a harmonic oscillator in the squeezed coherent state and the thermal squeezed coherent state. 
In Sec.~4. we demonstrate a simple extension of the dichotomic variable and the projection operator, which exhibits the violation of the Leggett-Garg inequality in the ground state and the squeezed state, which even boosts the violation. 
Sec.~5 is devoted to summary and conclusions.
In Appendix A, a brief review of deriving Eq.~(\ref{Hqsstt}) 
is presented. 
Similarly, in Appendix B, we derive a mathematical formula 
necessary to evaluate the quasi-probability distribution function for the thermal squeezed coherent state.
Throughout this paper, we use the unit $\hbar=1$.

\section{Basic Formulas}
In the first part of this section, we review the two-time Leggett-Garg inequalities with the two-time quasi-probability distribution function. 
We introduce a dichotomic variable $Q$, which gives $\pm1$ as a result of measurement. We define $Q_1$ and $Q_2$ to be the results of measurements at the time $t_1$ and $t_2$, respectively. 
Further, we introduce $s_1$ and $s_2$, which are to be chosen
$\pm1$ for the measurement at the time $t_1$ and $t_2$. 
Under these assumptions, the following inequalities hold
$(1+s_1Q_1)(1+s_2Q_2)\geq 0$ for the four combination of $s_1$ and $s_2$. 

Within a framework of macrorealism, there exists a probability function $p(Q_1,Q_2)$, which gives 
\begin{eqnarray}
&&\langle Q_1\rangle=\sum_{Q_1,Q_2}p(Q_1,Q_2)Q_1,~~~~
\langle Q_2\rangle=\sum_{Q_1,Q_2}p(Q_1,Q_2)Q_2,~~~~
\langle Q_1Q_2\rangle=\sum_{Q_1,Q_2}p(Q_1,Q_2)Q_1Q_2.
\end{eqnarray}
The probability function takes values of the range between $0$ and $1$, the expectation value of $(1+s_1Q_1)(1+s_2Q_2)$
must be non-negative.
\begin{eqnarray}
\langle(1+s_1Q_1)(1+s_2Q_2)\rangle\geq 0,
\end{eqnarray}
which is regarded as the Leggett-Garg inequalities of two-time. 

On the basis of the framework of quantum mechanics, introducing the dichotomic quantum variable $\hat Q$,
the corresponding variables $\hat Q_1$ and $\hat Q_2$ are defined by
$\hat Q_1={\hat Q}(t_1)=e^{iHt_1}\hat Qe^{-iHt_1}$ and
$\hat Q_2={\hat Q}(t_2)=e^{iHt_2}\hat Qe^{-iHt_2}$,
respectively, 
where we assume the unitary evolution of the system described by the   Hamiltonian operator $\hat H$. 
Then, the quasi-probability is defined by 
\begin{eqnarray}
&&q_{s_1,s_2}(t_1,t_2)=\frac{1}{8}{\rm Tr}[(1+s_1\hat Q_1)(1+s_2\hat Q_2)\rho_0]+(1\leftrightarrow2),
\end{eqnarray}
where $\rho_0$ is the density matrix of the initial state.
Introducing the Heisenberg operator by
\begin{eqnarray}
&&P_s(t)=e^{i\hat{H}t}P_se^{-i\hat{H}t}={\frac{1}{2}}e^{i\hat{H}t}(1+s\hat Q)e^{-i\hat{H}t}, 
\end{eqnarray}
where $P_s=(1+s\hat Q)/2$ is regarded as a projection operator, the quasi-probability is written as
\begin{eqnarray}
  &&q_{s_1,s_2}(t_1,t_2)={1\over 2}{\rm Tr}[P_{s_1}(t_1)P_{s_2}(t_2)\rho_0]+(1\leftrightarrow2)={\rm Re}
  {\rm Tr}[P_{s_2}(t_2)P_{s_1}(t_1)\rho_0].
\end{eqnarray}

We now apply the above formulation of the two-time quasi-probability distribution function for a continuous quantum variable of a harmonic oscillator. The same problem has been investigated in Refs.~\cite{Halliwell22,Halliwell23}. 
In the present paper, we present a new method to compute the quasi-probability distribution function, which is different from the previous work but reproduces the same results. 
We consider a harmonic oscillator described by the position $x$ with the angular frequency $\omega$ and 
the mass $m$.
When adopting the dichotomic variable $Q={\rm  sgn}(x)$, we have
\begin{eqnarray}
  P_s={1\over 2}(1+s\times{\rm sgn}(x))=\theta(sx),
\end{eqnarray}
where $\theta(x)$ is the Heaviside function.
Using the mathematical formula $\theta'(x-\alpha)=\delta(x-\alpha)={1\over 2\pi}\int_{-\infty}^\infty dp e^{-ip(x-\alpha)}$, i.e.,
\begin{eqnarray}
    -{d\over d\alpha}\theta(x-\alpha)={1\over 2\pi}\int_{-\infty}^\infty dp e^{-ip(x-\alpha)},
\end{eqnarray}
we have
\begin{eqnarray}
    \theta(x-\alpha)=\int^{\infty}_\alpha dc{1\over 2\pi}\int_{-\infty}^\infty dp e^{-ip(x-c)}.
\end{eqnarray}
Using the creation and annihilation operators, $ \hat a^\dagger$ and $ \hat a$, 
the position operator of $x$ is written as $
\hat x={(\hat a+{ \hat a}^\dagger)/ \sqrt{2m\omega}}$, and
we have
\begin{eqnarray}
\label{theta}
P_s(t)=e^{i\hat{H}t}\theta(s\hat{x})e^{-i\hat{H}t}=\int_0^\infty dc \int_{-\infty}^\infty {dp\over 2\pi} \exp\left[ip\left(-s{\hat a e^{-i\omega t}+\hat a^\dagger e^{i\omega t}\over\sqrt{2m\omega}}+c\right)\right].
\end{eqnarray}
Hence, the quasi-probability is expressed as
\begin{eqnarray}
  q_{s_1,s_2}(t_1,t_2)
  &=&{\rm Re~Tr}
\biggl[\iint_{0}^{\infty}dc_{1}dc_{2}\iint_{-\infty}^{\infty}\frac{dp_{1}dp_{2}}{(2\pi)^{2}}
  e^{-ip_{2}s_{2}(\hat{a}e^{-i\omega t_{2}}+\hat{a}^\dagger e^{i\omega t_{2}})+ip_{2}c_{2}}e^{-ip_{1}s_{1}(\hat{a}e^{-i\omega t_{1}}+\hat{a}^\dagger e^{i\omega t_{1}})+ip_{1}c_{1}}\rho_{0}\biggr],
\end{eqnarray}
where we redefined the integral variables as $p_i/\sqrt{2m\omega}\rightarrow p_i$ and $c_i\sqrt{2m\omega}\rightarrow c_i$ with $i=1,2$.

We next apply the above formula for the coherent squeezed state of a harmonic oscillator, 
which is defined by
\begin{eqnarray}
  |\xi,\zeta\rangle=D(\xi)S(\zeta)|0\rangle.
\end{eqnarray}
Here the squeezing operator $S(\zeta)$ and the displacement operator $D(\xi)$ are defined by $S(\zeta)=e^{{1\over 2}(\zeta \hat a^\dagger{}^2-\zeta^*\hat a^2)}$ and $D(\xi)=e^{\xi \hat a^\dagger-\xi^* \hat a}$, respectively, where $\xi$ and $\zeta$ are the parameters of complex number. We note that 
$S^\dagger(\zeta)=S(-\zeta)$ and $D^\dagger(\xi)=D(-\xi)$ .
By using the mathematical formula,
\begin{eqnarray}
 D(\xi)S(\zeta)=S(\zeta)D(\gamma),
 \label{SDrelation}
\end{eqnarray}
where 
\begin{eqnarray}
 \gamma=\xi\cosh |\zeta|-\xi^*e^{i\theta_0}\sinh |\zeta|
 \end{eqnarray}
with $ \zeta=|\zeta|e^{i\theta_0} $, 
the quasi-probability distribution function is written as
\begin{eqnarray}
  q_{s_1,s_2}(t_1,t_2)
  &=&
  {\rm Re}[ \langle 0|D^\dagger(\gamma)S^\dagger(\zeta) e^{i\hat H t_2}\theta(s_2\hat x)e^{-i\hat H t_2}e^{i\hat H t_1}\theta(s_1\hat x)e^{-i\hat H t_1}S(\zeta)D(\gamma)|0\rangle].
\end{eqnarray}
Further, introducing the operator of the Bogoliubov transformation, $\hat b$ and $\hat b^\dagger$, 
\begin{eqnarray}
&&S^\dagger(\zeta)\hat{a}S(\zeta)=\hat{a}\cosh r+\hat{a}^\dagger e^{i\theta_0}\sinh r=\hat{b},\\
&&S^\dagger(\zeta)\hat{a}^\dagger S(\zeta)=\hat{a}^\dagger \cosh r+\hat{a}e^{-i\theta_0}\sinh r=\hat{b}^\dagger,
\end{eqnarray}
where we used $r=|\zeta|$, we have
\begin{eqnarray}
&&S^\dagger(\zeta) e^{i\hat{H}t_2}\theta(s_2\hat x)e^{-i\hat{H}t_2}e^{i\hat{H}t_1}\theta(s_1\hat{x})e^{-i\hat{H}t_1}S(\zeta)
\nonumber\\
&&~~~~=\int_0^\infty dc_1 e^{ip_1c_1}\int_{-\infty}^\infty {dp_1\over 2\pi}
\int_0^\infty dc_2 e^{ip_2c_2}\int_{-\infty}^\infty {dp_2\over 2\pi}
 e^{-ip_2s_2(\hat{b}e^{-i\omega t_2}+\hat{b}^\dagger e^{i\omega t_2})}
e^{-ip_1s_1(\hat{b}e^{-i\omega t_1}+\hat{b}^\dagger e^{i\omega t_1})}
\end{eqnarray}
 and 
\begin{eqnarray}
&&\langle0|D^\dagger(\gamma)S^\dagger(\zeta) e^{i\hat{H}t_2}\theta(s_2\hat{x})e^{-i\hat{H}t_2}e^{i\hat{H}t_1}\theta(s_1\hat{x})e^{-i\hat{H}t_1}S(\zeta)D(\gamma)|0\rangle
\nonumber\\
&&~~=\int_0^\infty dc_1 \int_0^\infty dc_2 \int_{-\infty}^\infty {dp_1\over 2\pi}
\int_{-\infty}^\infty {dp_2\over 2\pi}e^{ip_1c_1}e^{ip_2c_2}
\langle0|D^\dagger(\gamma)e^{-ip_2s_2(E(t_2)\hat{a}+E^*(t_2)\hat{a}^\dagger)}
e^{-ip_1s_1(E(t_1)\hat{a}+E^*(t_1)\hat{a}^\dagger)}D(\gamma)|0\rangle
\nonumber\\
\end{eqnarray}
with defined
$E(t)=e^{-i\omega t}\cosh r +e^{i\omega t}e^{-i\theta_0}\sinh r $.

Repeatedly using the BCH formula, $e^{A+B}=e^A e^B e^{-[A,B]/2}$ and $e^A e^B=e^{[A,B]}e^Be^A$, which hold for the operators $A$ and $B$ satisfying $[A,B]={\rm constant}$, we have 
\begin{eqnarray}
&&\langle0|D^\dagger(\gamma)S^\dagger(\zeta) e^{i\hat{H}t_2}\theta(s_2\hat{x})e^{-i\hat{x}t_2}e^{i\hat{x}t_1}\theta(s_1\hat{x})e^{-i\hat{x}t_1}S(\zeta)D(\gamma)|0\rangle
\nonumber\\
&&~~=\int_0^\infty dc_1 \int_0^\infty dc_2 \int_{-\infty}^\infty {dp_1\over 2\pi}
\int_{-\infty}^\infty {dp_2\over 2\pi}e^{ip_1c_1}e^{ip_2c_2}
e^{-ip_1s_1(E(t_1)\gamma+E^*(t_1)\gamma^*)-ip_2s_2(E(t_2)\gamma+E^*(t_2)\gamma^*)}
\nonumber\\
&&~~~~~~
\exp\left[-{1\over 2}\left(|E(t_2)|^2p_2^2+|E(t_1)|^2 p_1^2+2p_1p_2s_1s_2 E(t_2) E^*(t_1)\right)\right].
\label{DSTSD}
\end{eqnarray}
The integration in Eq.~(\ref{DSTSD}) with respect to $p_1$ and $p_2$ can be performed as
\begin{eqnarray}
  \int_{-\infty}^\infty dp_1\int_{-\infty}^\infty dp_2
  \exp\left[-{1\over 2}\bf p^T\bm A\bm p+\bm \rho^T \cdot\bm p\right]={2\pi\over \sqrt{{\rm det}\bm A}}
  \exp\left[{1\over 2}\bm \rho^T\bm A^{-1}\bm\rho\right],
\end{eqnarray}
where $\bm A$ and $\rho$ are read
\begin{eqnarray}
&&\bm A=\left(
\begin{array}{cc}
|E(t_1)|^2&s_1s_2E(t_2)E^*(t_1)\\
s_1s_2E(t_2)E^*(t_1)&|E(t_2)|^2
\end{array}\right)
,\\
&&\bm \rho^T=\Bigl(ic_1-is_1(E(t_1)\gamma+E^*(t_1)\gamma^*),ic_2-is_2(E(t_2)\gamma+E^*(t_2)\gamma^*)\Bigr).
  \end{eqnarray}
Further, the integration with respect to $c_1$ and $c_2$
can be written by setting $c_1=c\cos u$ and $c_2=c\sin u$, and we have
\begin{eqnarray}
&&\langle0|D^\dagger(\gamma)S^\dagger(\zeta) e^{iHt_2}\theta(s_2\hat x)e^{-iHt_2}e^{iHt_1}\theta(s_1\hat x)e^{-iHt_1}S(\zeta)D(\gamma)|0\rangle
={1\over 2\pi}\int_0^{\pi/2}du\int_0^\infty dc c {1\over \sqrt{ B}} 
e^{-{1\over 2}C},
\end{eqnarray}
where we defined 
\begin{eqnarray}
&&B=\det \bm A=|E(t_1)|^2|E(t_2)|^2-(E^*(t_1)E(t_2))^2,
\\
&&C={1\over B}\biggl(|E(t_2)|^2(c\cos u+s_1{{\cal E}(t_1)})^2+|E(t_1)|^2(c\sin u+s_2{\cal E}(t_2))^2
\nonumber\\
&&~~~~~~-2s_1s_2E^*(t_1)E(t_2)(c\cos u-s_1{\cal E}(t_1))(c\sin u-s_2{\cal E}(t_2))\biggr),
\\
&&{\cal E}(t)=E(t)\gamma+E^*(t)\gamma^*.
\end{eqnarray}
By introducing the quantities,
\begin{eqnarray}
&&\sigma={1\over B}\Bigl(|E(t_2)|^2\cos^2 u+|E(t_1)|^2\sin^2 u-2s_1s_2E(t_2)E^*(t_1)\sin u\cos u
\Bigr),
\\
&&\beta={1\over B}\Bigl(-|E(t_2)|^2s_1{\cal E}(t_1)\cos u-|E(t_1)|^2s_2{\cal E}(t_2)\sin u+E(t_2)E^*(t_1)(s_2{\cal E}(t_1)\sin u+s_1{\cal E}(t_2)\cos u)
\Bigr),
\\
&&\delta={1\over B}\Bigl(|E(t_2)|^2{\cal E}(t_1)^2+|E(t_1)|^2{\cal E}(t_2)^2-2E(t_2)E^*(t_1){\cal E}(t_1){\cal E}(t_2),
\Bigr),
\end{eqnarray}
$C$ can be written as $C=\sigma c^2+2\beta c+\delta$, and we finally have
\begin{eqnarray}
  q_{s_1,s_2}(t_1,t_2)
  &=&
  {\rm Re}\left[ 
\langle0|D^\dagger(\gamma)S^\dagger(\zeta) e^{iHt_2}\theta(s_2\hat x)e^{-iHt_2}e^{iHt_1}\theta(s_1\hat x)e^{-iHt_1}S(\zeta)D(\gamma)|0\rangle\right]
\nonumber\\
&=&
{\rm Re}\left[{1\over 2\pi}{e^{-\delta/2}\over \sqrt{B}} \int_0^{\pi/2}du
\biggl\{{1\over \sigma}-{\sqrt{2\pi}be^{\beta^2/2\sigma}\over 2\sigma^{3/2}}
{\rm erfc}\left({\beta\over \sqrt{2\sigma}}\right)\biggr\}\right],
\label{qsstt}
\end{eqnarray}
where we used the mathematical formula
\begin{eqnarray}
\int_0^\infty dc c e^{-(\sigma c^2+2\beta \sigma+\delta)/2}=e^{-\delta/2}\biggl\{{1\over \sigma}-{\sqrt{\pi/2}\beta e^{\beta^2/2\sigma}\over 2\sigma^{3/2}}
{\rm erfc}\left({\beta\over \sqrt{2\sigma}}\right)\biggr\}, 
\end{eqnarray}
where ${\rm erfc}(z)$ is the complementary error function. 
As we will show in the next section, the integration can be evaluated numerically using the software Mathematica.

At the end of this section, we remark on a useful generalization of the above formula.
We may adopt the following projection operator by constructing
the dichotomic variable as $Q={\rm sgn}(\hat x-\bar x(t))$ instead of $Q={\rm sgn}(\hat x)$ 
\begin{eqnarray}
P_s={1\over 2}(1+s \times {\rm sgn}(\hat x-\bar x(t)))=\theta(s(\hat x-\bar x(t)))
=\int_0^\infty dc \int_{-\infty}^\infty {dp\over 2\pi} e^{ip(-s(\hat x-\bar x(t))+c)},
\label{sxb}
\end{eqnarray}
where $\bar x(t)$ is an arbitrary function of time. 
In this case, we have
\begin{eqnarray}
&&\langle0|D^\dagger(\gamma)S^\dagger(\zeta) e^{iHt_2}\theta(s_2(\hat x-\bar x(t_2)))e^{-iHt_2}e^{iHt_1}\theta(s_1(\hat x-\bar x(t_1)))e^{-iHt_1}S(\zeta)D(\gamma)|0\rangle
\nonumber\\
&&~~=\int_0^\infty dc_1 \int_0^\infty dc_2 \int_{-\infty}^\infty {dp_1\over 2\pi}
\int_{-\infty}^\infty {dp_2\over 2\pi}e^{ip_1c_1}e^{ip_2c_2}
e^{-ip_1s_1(E(t_1)\gamma+E^*(t_1)\gamma^*-\bar x(t_1))-ip_2s_2(E(t_2)\gamma+E^*(t_2)\gamma^*-\bar x(t_2))}
\nonumber\\
&&~~~~~~
\exp\left[-{1\over 2}\left(|E(t_2)|^2p_2^2+|E(t_1)|^2 p_1^2+2p_1p_2s_1s_2 E(t_2) E^*(t_1)\right)\right].
\label{DSTSDX}
\end{eqnarray}
Comparing this expression (\ref{DSTSDX}) with Eq.~(\ref{DSTSD}), we find that (\ref{DSTSDX}) is reproduced from (\ref{DSTSD}) with replacing $E(t)\gamma+E^*(t)\gamma^*$ with $E(t)\gamma+E^*(t)\gamma^*-\bar x(t)$.
This leads to an important implication. For example,
when the state of the harmonic oscillator is in the coherent state, i.e., $\zeta=0$, we have
$E(t)\gamma+E^*(t)\gamma^*=2|\xi|\cos(\omega t-\Theta)$, where we assumed $\xi=|\xi|e^{i\Theta}$. 
Therefore, the same quasi-probability distribution function is predicted between the case taking the coherent state with $\xi=|\xi|e^{i\Theta}$ and $\bar x(t)=0$ and the case taking the ground state but with $\bar x(t)=-2|\xi|\cos(\omega t-\Theta)$.
This means that it is possible to observe the violation of the Leggett-Garg inequalities in the system of a harmonic oscillator in the ground state by choosing $\bar x(t)$ properly. This also makes it possible to observe a similar violation of the inequalities in a harmonic oscillator in a squeezed state, although there appears no violation for the harmonic oscillator in the ground state or the squeezed state when we adopt $\hat Q={\rm sgn}(\hat x)$, i.e., $\bar x(t)=0$ in the above formulas.

\section{Mawby-Halliwell's formulation for squeezed coherent state}
\def\alphax{\xi}
\subsection{Squeezed coherent state}
In this section, we evaluate the quasi-probability distribution function following the method developed in the paper by Mawby and Halliwell \cite{Halliwell23}, in order to compare the results with those developed in the previous section. We first consider the case of the squeezed coherent 
state, which is written as
$    \Ket{\alphax,\zeta}=D(\alphax)S(\zeta)\Ket{0}
$
with the squeezing operator $S(\zeta)$ defined by $S(\zeta)=e^{\frac{1}{2}(\zeta \hat{a}^{\dagger2}-\zeta^{*}\hat{a}^2)}$ and the coherent operator $D(\alphax)$ defined by $D(\alphax)=e^{\alphax\hat{a}^{\dagger}-\alphax^{*}\hat{a}}$, where 
we use $ \zeta=r e^{i\theta_0} $.
An explicit expression for the quasi-probability for the squeezed coherent state, 
which was not given in \cite{Halliwell23}, has the 
following form.  Its derivation is shown in Appendix A;
\begin{eqnarray}
    q_{s_1,s_2}(t_1,t_2)&=&\text{Re}\biggl[\sum^{\infty}_{n=0}e^{-in\omega(t_2-t_1)-in(\beta(t_2)-\beta(t_1))}\Bra{0}\theta\left(s_2(\hat{x}+\frac{x_{\alphax(t_2)}}{\lambda(t_2)})\right)\Ket{n}\Bra{n}\theta\left(s_1(\hat{x}+\frac{x_{\alphax(t_1)}}{\lambda(t_1)})\right)\Ket{0}\biggr]
\nonumber\\
&=&\frac{1}{4}\left\{1+s_1\mathrm{erf}\left(\frac{x_{\alphax(t_1)}}{\lambda(t_1)}\right)+s_2\mathrm{erf}\left(\frac{x_{\alphax(t_2)}}{\lambda(t_2)}\right)+s_1s_2\mathrm{erf}\left(\frac{x_{\alphax(t_1)}}{\lambda(t_1)}\right)\mathrm{erf}\left(\frac{x_{\alphax(t_2)}}{\lambda(t_2)}\right)\right\}\nonumber\\
&&\hspace{20pt}+s_1s_2\text{Re}\sum^{\infty}_{n=1}e^{-in\omega(t_2-t_1)-in(\beta(t_2)-\beta(t_1))}J_{0n}\left(-\frac{x_{\alphax(t_1)}}
{\lambda(t_1)},\infty\right)J_{n0}\left(-\frac{x_{\alphax(t_2)}}{\lambda(t_2)},\infty\right),
\label{Hqsstt}
\end{eqnarray}
where $\lambda(t)$ is 
\begin{eqnarray}
\lambda(t)={\sqrt{\sinh (2 r) \cos (2\omega t-\theta_0)+\cosh (2 r)}},
\end{eqnarray}
and $J_{mn}(x_1,x_2)$ is defined as the matrix element 
in Ref.~\cite{Halliwell22} as
\begin{eqnarray}
&& J_{mn}(x_1,x_2)=\int^{x_2}_{x_1}dx\braket{m|x}\braket{x|n},
\end{eqnarray}
where $|m\rangle$ and $|n\rangle$ are the energy eigenstates of the harmonic oscillator with the non-negative integers $m$ and $n$.
For $m\neq n$, $J_{mn}(x_1,x_2)$ 
is given by
\begin{eqnarray}
    J_{mn}(x_1,x_2)=\frac{1}{2(\epsilon_{n}-\epsilon_{m})}[\psi^{\prime}_m(x_2)\psi_n(x_2)-\psi^{\prime}_n(x_2)\psi_m(x_2)-\psi^{\prime}_m(x_1)\psi_n(x_1)+\psi^{\prime}_n(x_1)\psi_m(x_1)],
\end{eqnarray}
where $\psi_j(x)$  and  $\epsilon_j$ with $j=0,1,2,\cdots$ are the energy eigenfunction $\psi_j(x)=\langle x|j\rangle$ and the corresponding energy eigenvalue, respectively, and   
the prime denotes the differentiation w.r.t the argument, i.e., $\psi^\prime(x)=d\psi(x)/dx$. 
One can use the following formulas 
\begin{eqnarray}
    J_{mn}(x,\infty)=\frac{1}{2(\epsilon_{n}-\epsilon_{m})}[-\psi^{\prime}_m(x)\psi_n(x)+\psi^{\prime}_n(x)\psi_m(x)]
\end{eqnarray}
in the expression (\ref{Hqsstt}), and 
\begin{eqnarray}
    J_{00}(x,\infty)=\frac{1}{2}(1-\mathrm{erf}(x)),
\end{eqnarray}
where $\erf(z)$ is the error function.

\begin{figure}[tbp]
       \centering
\includegraphics[width=15cm]{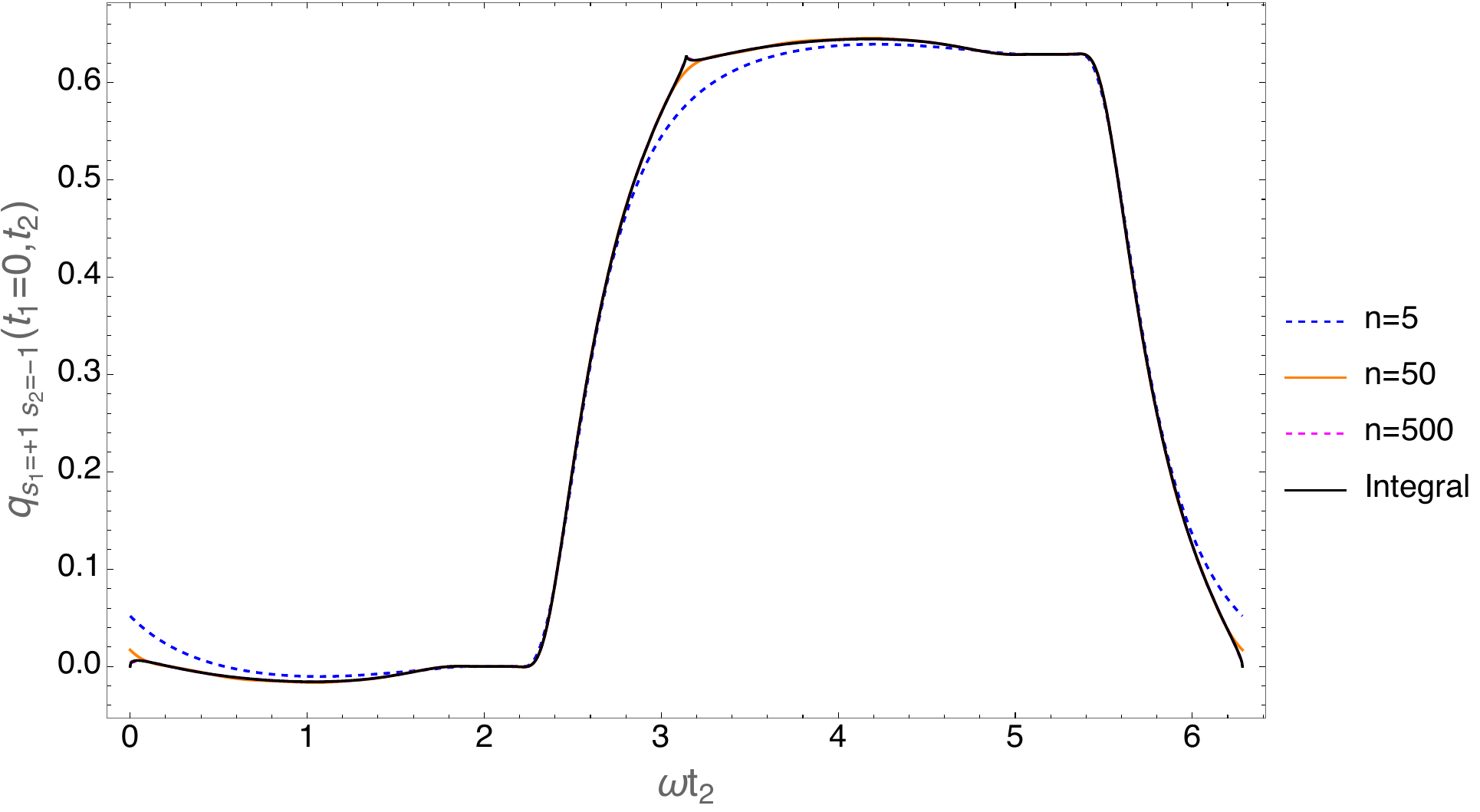}
\caption{Comparison of the numerical results of Eq.~(\ref{qsstt}) and Eq.~(\ref{Hqsstt}). In the numerical evaluation of Eq.~(\ref{Hqsstt}), we increased the maximum number of the sum with respect to $n$ up to $n_{}=5$ (blue dashed curve), $n_{}=50$ (orange solid curve), $n_{}=500$ (red dashed curve), where the red dashed curve is not seen in the figure. 
The black solid curve plots the numerical integration of Eq.~(\ref{qsstt}), which overlaps the red dashed curve, i.e., Eq.~(\ref{Hqsstt}) with $n_{}=500$. 
Thus, the two formulas Eq.~(\ref{qsstt}) and Eq.~(\ref{Hqsstt})
coincide as long as the sum with respect to $n$ is taken sufficiently large. 
Here we fixed $x_0=0.550$, $p_0=1.925$, $r=1$, $\theta_0={\pi}/{3}$, $s_1=1$, and $s_2=-1$. 
\label{figone}
}
\end{figure}

We define the parameter for the initial state of the coherent state by
$    x_0=\sqrt{2}{\rm Re}[\xi]=\sqrt{2}|\xi|\cos\Theta$,
$    p_0=\sqrt{2}{\rm Im}[\xi]=\sqrt{2}|\xi|\sin\Theta$.
Here we note that $x_0$ and $p_0$ are understood as the initial 
values of the position and momentum of coherent oscillating motion, normalized by the factors $1/\sqrt{2m\omega}$
and $\sqrt{m\omega/2}$, respectively.
Figure \ref{figone} shows how the
two formulas Eq.~(\ref{qsstt}) and Eq.~(\ref{Hqsstt}) give the same result, 
which plots  
the quasi-probability distribution function $q_{1,-1}(0,t_2)$ as a function of $t_2$, 
where we fixed $t_1=0$, $x_0=0.550$,~$p_0=1.925$, $s_1=1$, $s_2=-1$, $r=1$, and $\theta_0=\pi/3$. 
Fig.~\ref{figone} demonstrates that
the numerical sum of Eq.~(\ref{Hqsstt}) 
and the numerical integration of Eq.~(\ref{qsstt}) converge
as the maximum value of the sum with respect to $n$ in 
Eq.~(\ref{Hqsstt}) increases. 
Thus, Fig.~\ref{figone} demonstrates that the same result is obtained from the two different formulas Eq.~(\ref{qsstt}) and Eq.~(\ref{Hqsstt}).

\begin{figure}[htbp]
    \begin{tabular}{cc}
          \begin{minipage}[t]{0.5\hsize}
        \centering
\includegraphics[keepaspectratio, scale=0.48]{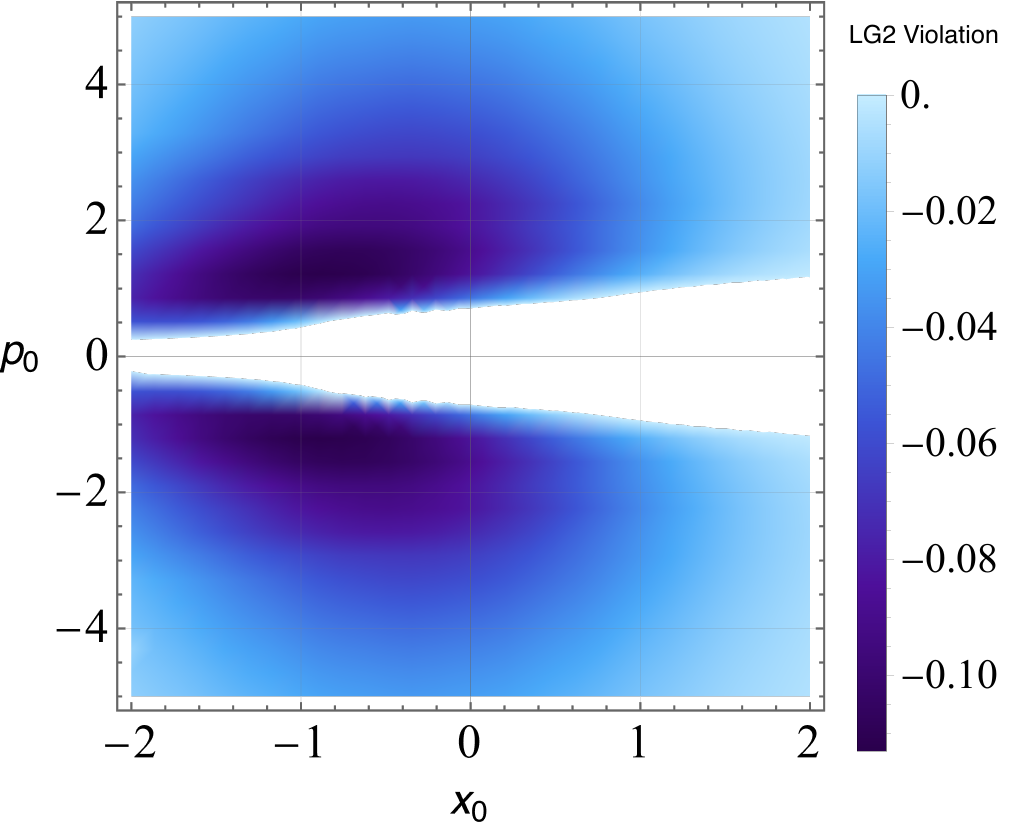}
        \subcaption{$s_1=+1, s_2=-1$}
      \end{minipage} &
      \begin{minipage}[t]{0.5\hsize}
      \centering
\includegraphics[keepaspectratio, scale=0.48]{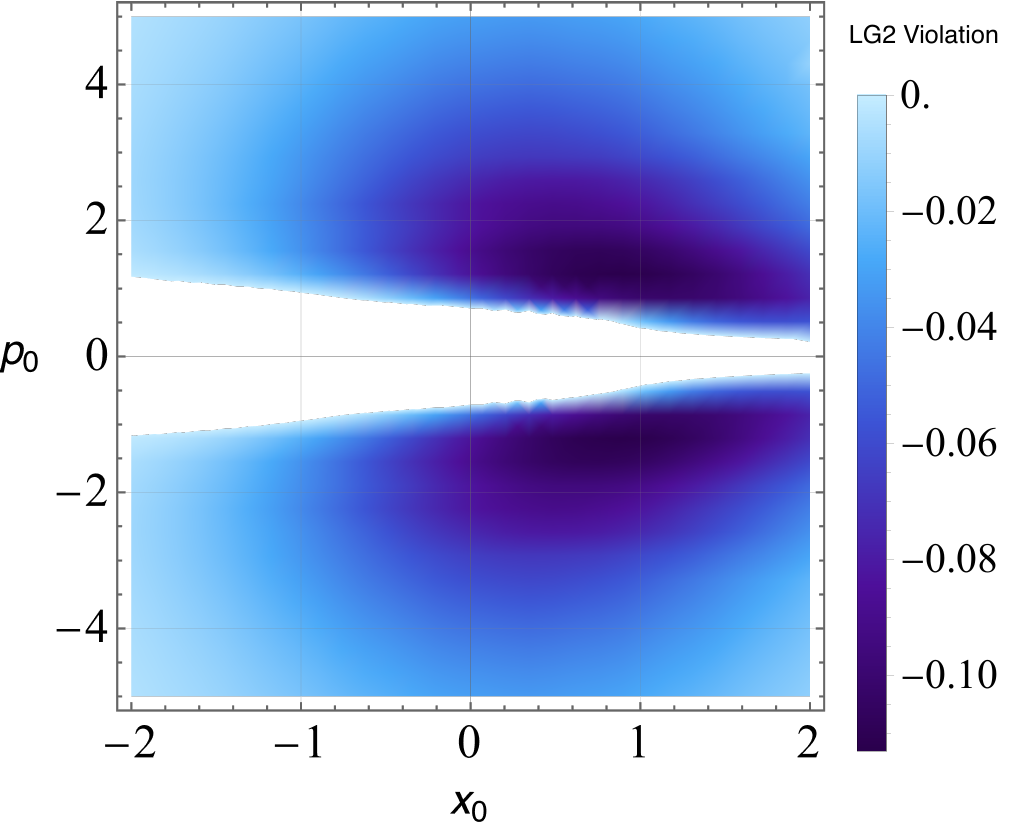}
        \subcaption{$s_1=-1, s_2=+1$}
      \end{minipage} 
      \vspace{1cm}
      \\
     \begin{minipage}[t]{0.5\hsize}
        \centering
        \includegraphics[keepaspectratio, scale=0.48]{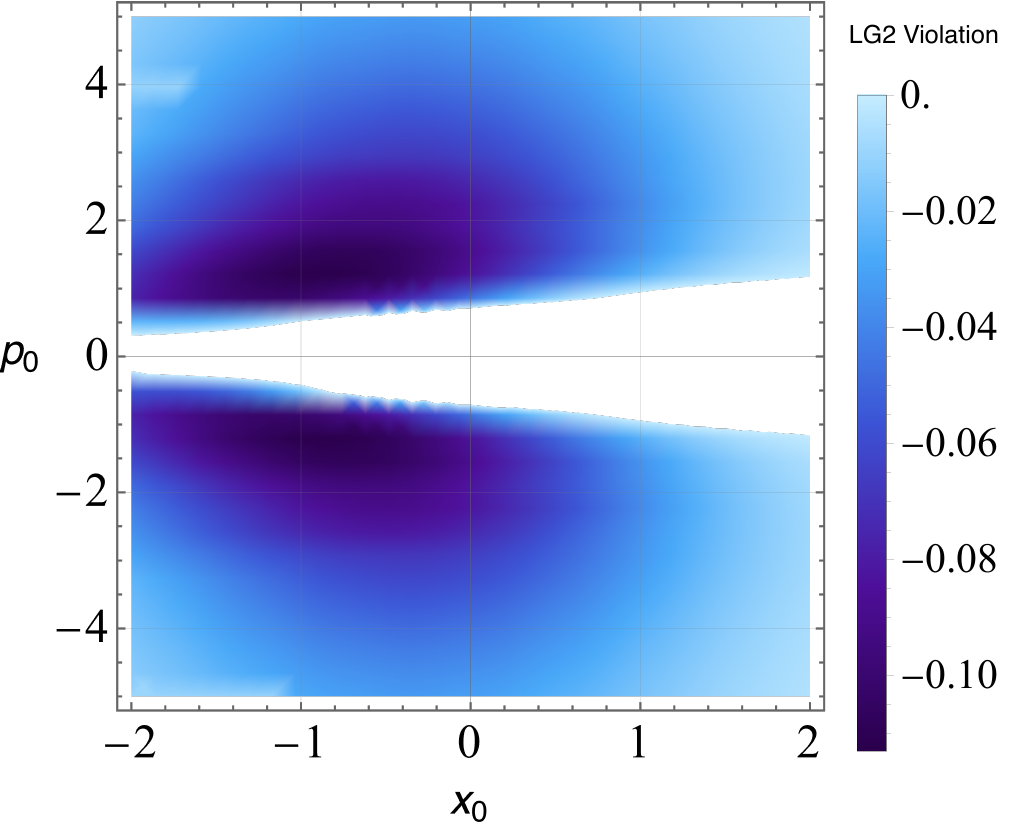}
        \subcaption{$s_1=+1, s_2=+1$}
      \end{minipage} &
      \begin{minipage}[t]{0.5\hsize}
        \centering
        \includegraphics[keepaspectratio, scale=0.48]{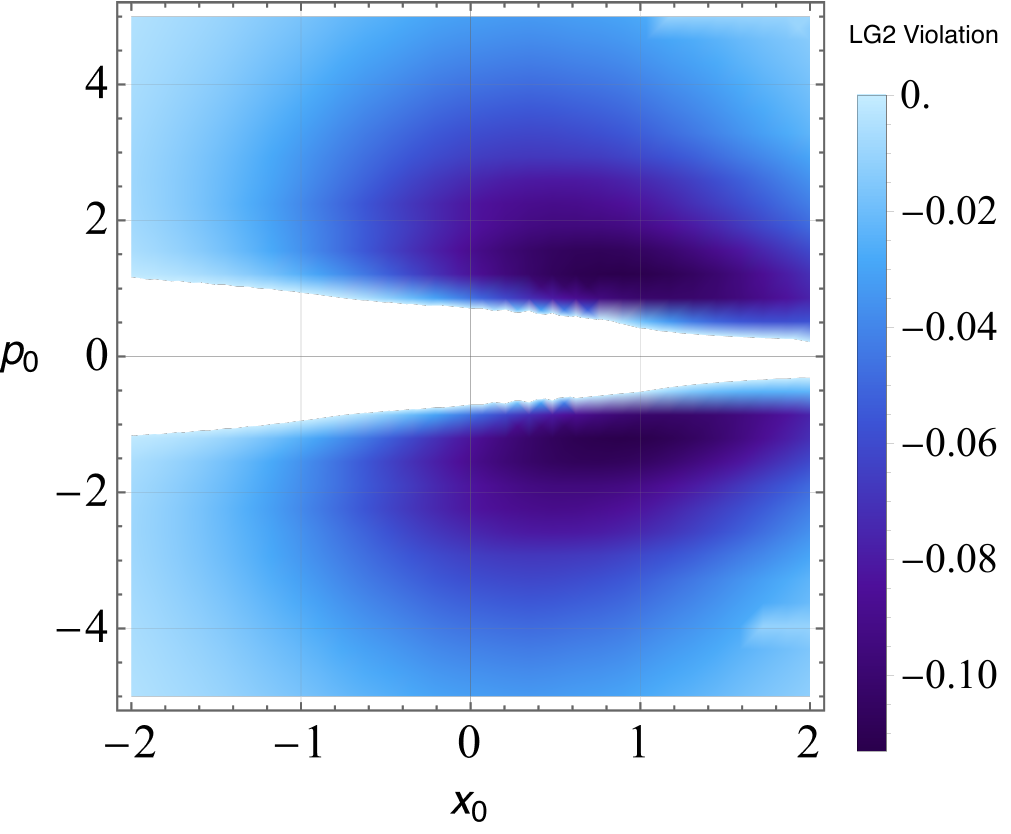}
        \subcaption{$s_1=-1, s_2=-1$}
      \end{minipage} 
    \end{tabular}
     \caption{ Contour plot of the minimum values of the two-time quasi-probability distribution function multiplied by the factor four, $4q_{s_1,s_2}(0,t_2)$, on the plane of $x_0$ and $p_0$.
     Each panel assumes (a) $s_1=1, s_2=-1$,  (b) $s_1=-1, s_2=1$,
     (c) $s_1=1, s_2=1$,  and (d) $s_1=-1, s_2=-1$. 
     In each panel, we fixed the squeezing parameter
     $r=1/2$ and $\theta_0=0$.\label{pxpx}}
  \end{figure}

We now consider the minimum value of the quasi-probability distribution function for the squeezed coherent state. 
Similar analysis was done for the coherent state in Ref.~\cite{Halliwell23}.
Following Ref.~\cite{Halliwell23}, figure \ref{pxpx} plots the minimum value of the two-time quasi-probability distribution function multiplied by the factor four.
Each panel of Fig.~\ref{pxpx} shows the minimum values of the two-time quasi-probability distribution function multiplied by $4$ on the plane of $x_0$ and $p_0$,
where $s_1$ and $s_2$ are fixed as 
$s_1=1$ and $s_2=-1$ (upper left panel), 
$s_1=-1$ and $s_2=1$ (upper right panel), $s_1=1$ and $s_2=1$ (lower left panel), $s_1=-1$ and $s_2=-1$ (lower right panel). 
In these panels, we adopted $r=1/2$ and $\theta_0=0$.
In each panel, the Leggett-Garg inequalities are violated in the colored regions, but they are not violated in the white regions. 
The quasi-probability distribution function takes smaller negative values in the darker blue regions, where the violation is larger than in the lighter blue regions. 
The smallest value in each panel in Fig. \ref{pxpx} is $-0.113$, which is the same for all the panels. Thus the minimum value of the two-time quasi-probability distribution function multiplied by the factor four is $-0.113$, which is the  $22.6\%$  of the L\"{u}der's Bound $-1/2(=4\times(-1/8))$, the maximum violation \cite{Budroni,Halliwell22}. 
This minimum value is the same as that for the coherent states 
investigated in Ref.~\cite{Halliwell22}, demonstrating that squeezing has no impact on boosting the magnitude of the violation of the Leggett-Garg inequalities in our model.
This is explicitly demonstrated in an analytic manner in the below (see also Ref.~\cite{Halliwell22}).

\begin{figure}[t]
   \begin{center}
\includegraphics[width=85mm]{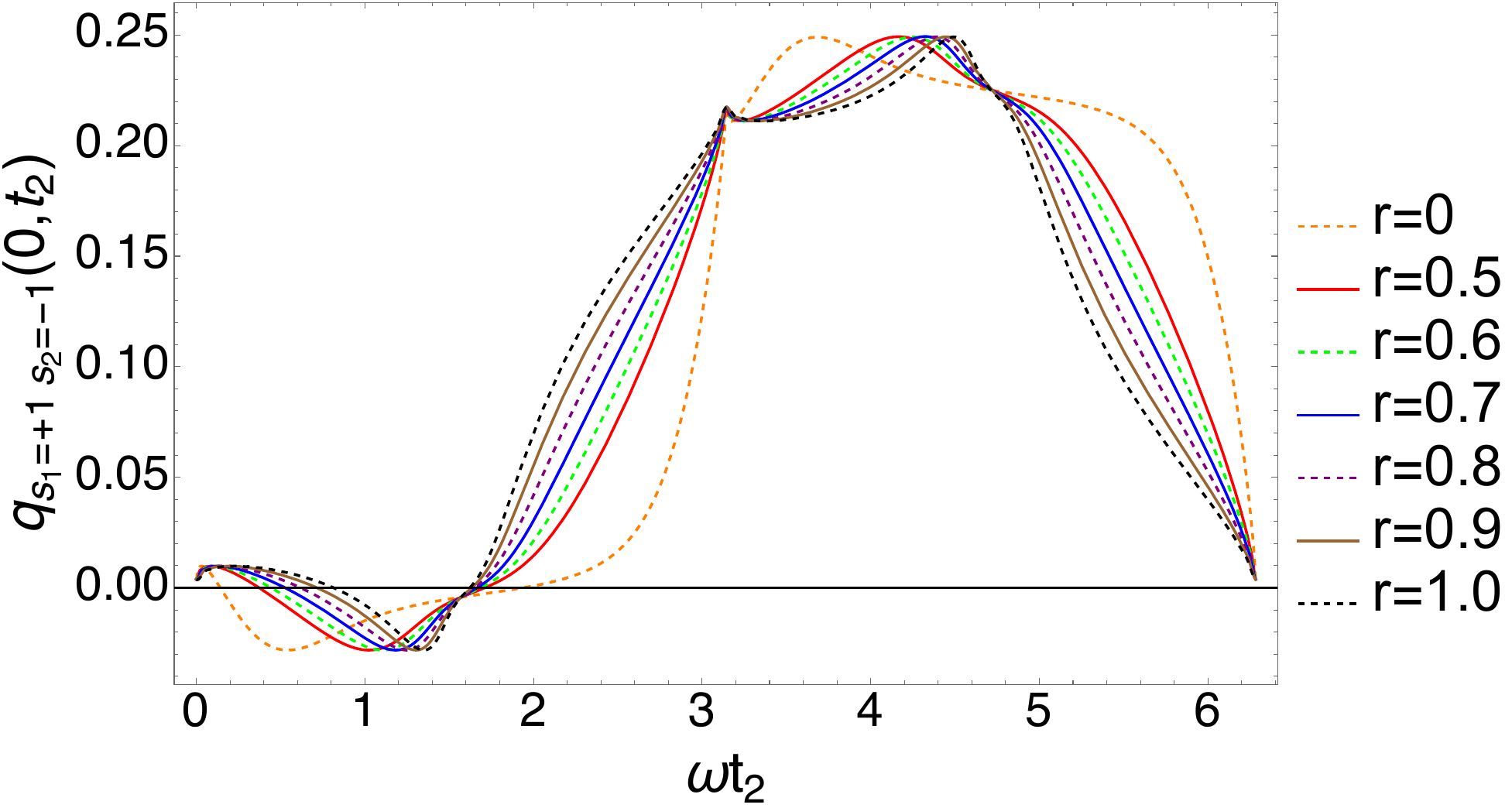}
   \hspace{.5cm}
\includegraphics[width=85mm]{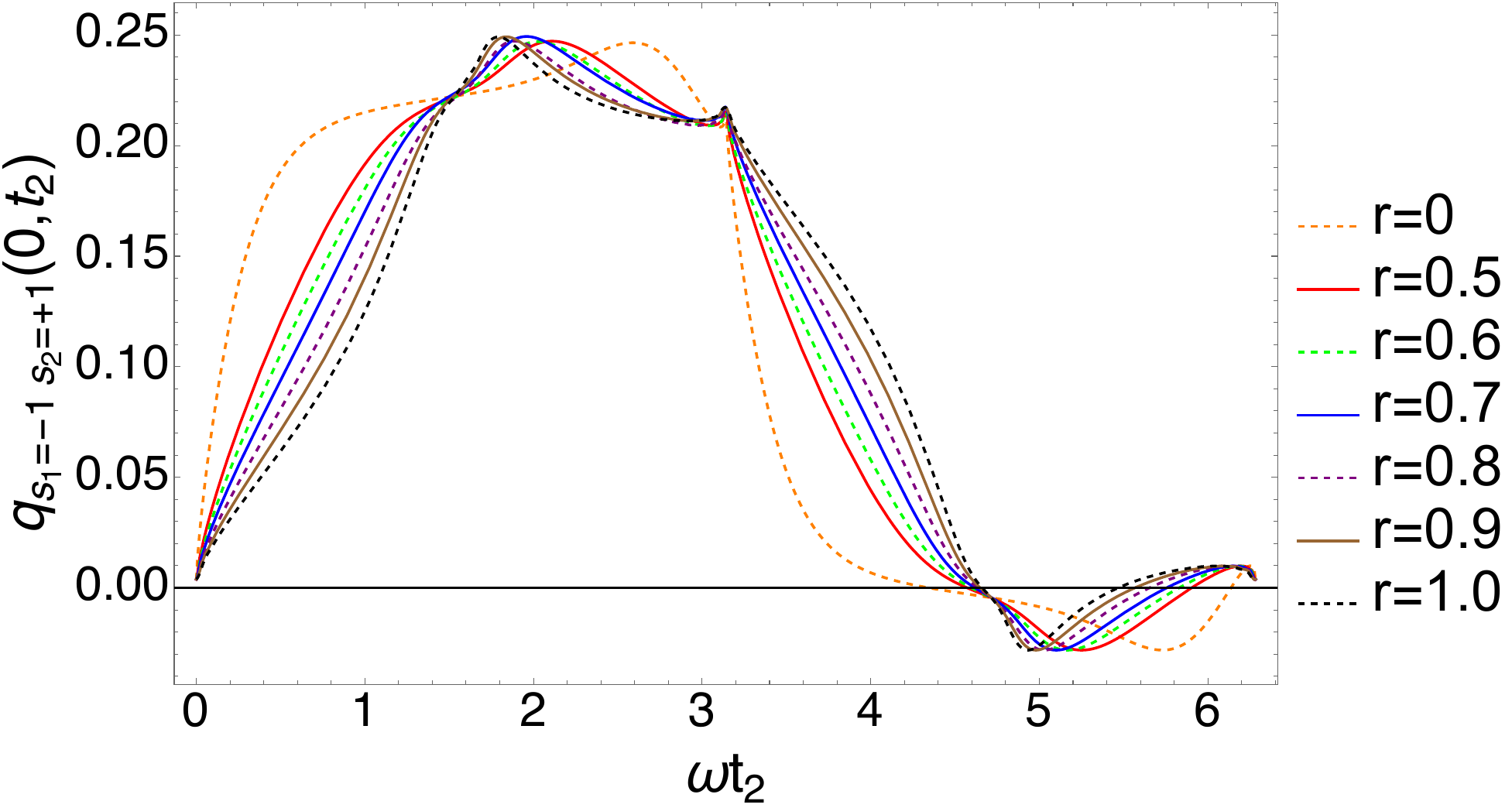}
     \caption{$q_{s_1,s_2}(0,t_2)$ as function of $\omega t_2$, which achieves the minimum value $4q_{s_1,s_2}(0,t_2)=-0.113$. 
     The left panel adopted $s_1=1$,~$s_2=-1$, while the right panel adopted $s_1=-1$,~$s_2=1$. 
     In both the panels we fixed $\theta_0=0$, and the other parameters are noted in the Table I.}
      \label{Demomin}
  \end{center}
\vspace{1cm}
   \begin{center}
\includegraphics[width=85mm]{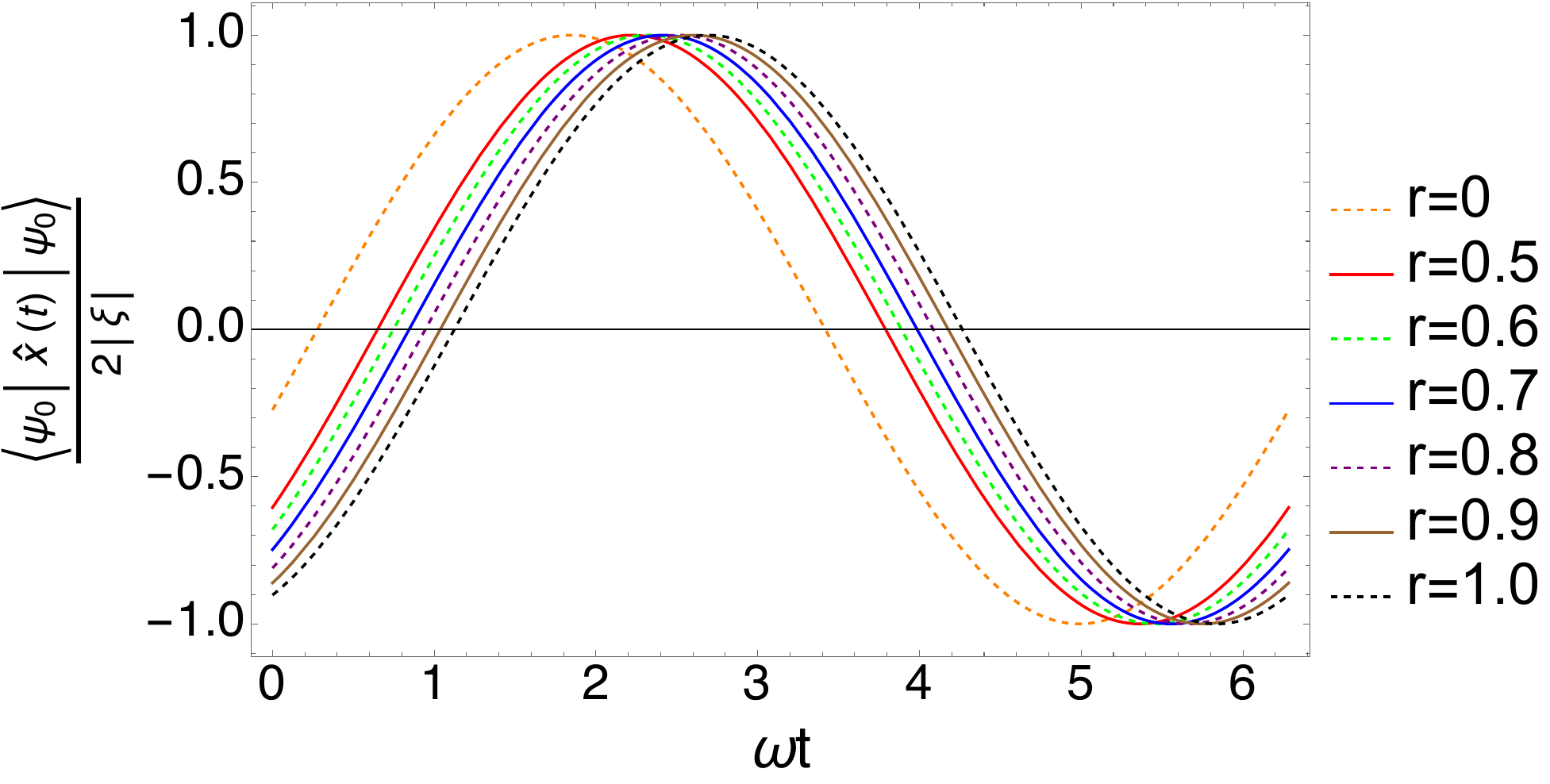}
   \hspace{0.5cm}
\includegraphics[width=85mm]{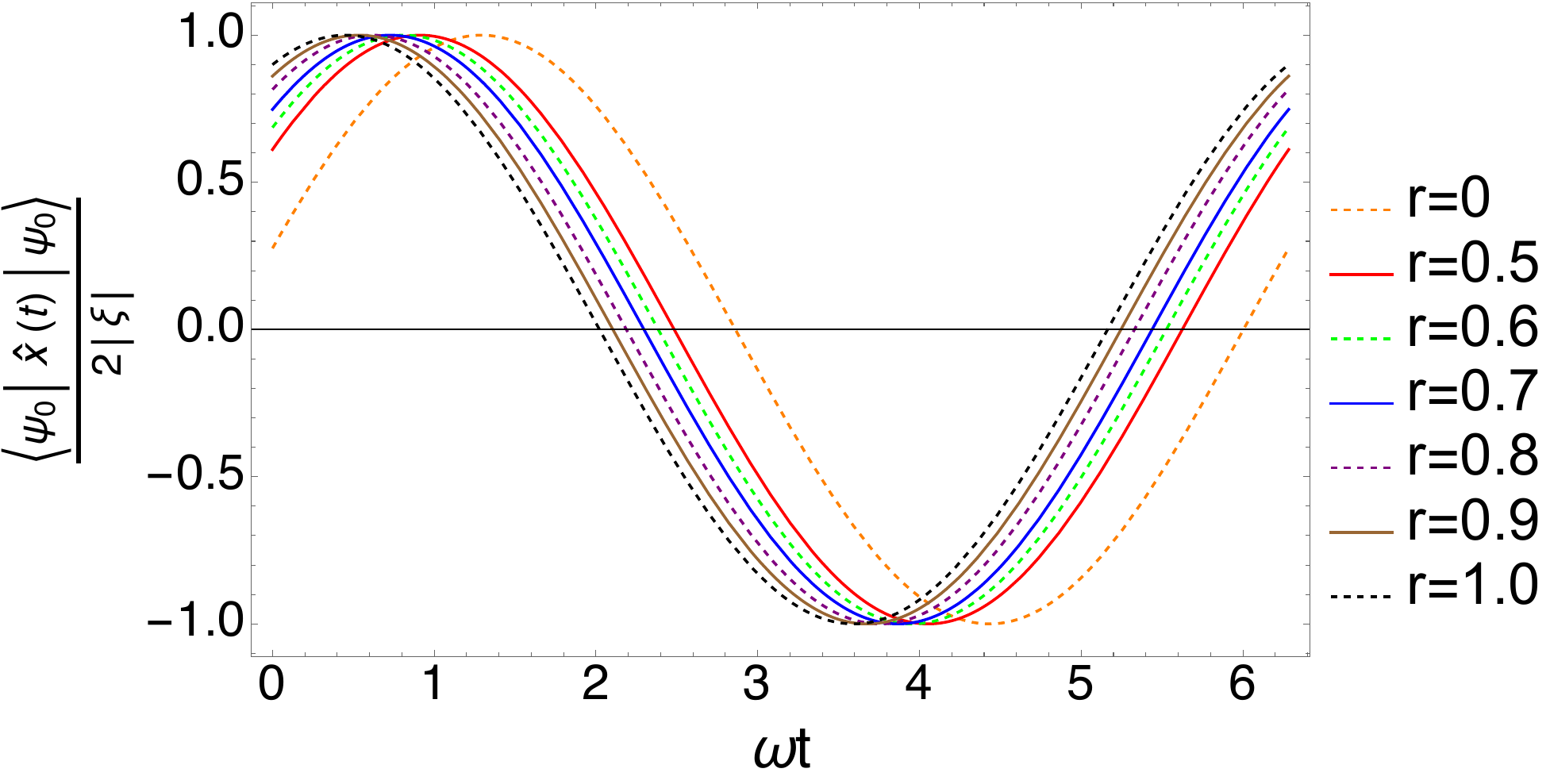}
     \caption{
     $\langle\psi_0|\hat x(t)|\psi_0\rangle/2|\xi|=\cos(\omega t-{\rm ArcTan}[p_0/x_0])$
     as a function of $\omega t$.
     The same-type curves in the left (right) panel and in the left (right) panel of Fig.~\ref{Demomin} assume the same parameters for each described in Table I.
     }
      \label{HO}
  \end{center}
  \end{figure} 
\begin{table}[btp]
   \caption{Parameters adopted for the curves in Figure \ref{Demomin}.}
   \label{parameters}
   \centering
    \begin{tabular}{||cclll||ccclll||}
\hline
~~$s_1$~~ & ~~~~$s_2$~~~~~ & ~~$r$~~~~~~ & ~~~$x_0$~~~~~~~ & ~~$p_0$~~~~~ &  & ~~$s_1$~ & ~~~~~$s_2$~~~~~ & ~~$r$~~~~~~ & ~~~$x_0$~~~~~~~ & ~~$p_0$~~~~~ \\ \hline 
1 & -1 & 0 & -0.554 & 1.95 &  & -1 & 1 & 0 & 0.550 & 1.93 \\ \hline
1 & -1& 0.5 & -0.896 & 1.18 &  & -1 & 1& 0.5 & 0.904 & 1.17 \\ \hline
1 & -1 & 0.6 & -0.991 & 1.07 &  & -1 & 1 & 0.6 & 1.00 & 1.06 \\ \hline
1 & -1 & 0.7 & -1.09 & 0.968 &  & -1 & 1 & 0.7 & 1.09 & 0.968 \\ \hline
1 & -1  & 0.8 & -1.21 & 0.875 &  & -1 & 1 & 0.8 & 1.22 & 0.866 \\ \hline
1 & -1 & 0.9 & -1.34 & 0.792 &  & -1 & 1 & 0.9 & 1.34 & 0.792 \\ \hline
1 & -1 & 1.0 & -1.48 & 0.717 &  & -1 & 1 & 1.0 & 1.48 & 0.717 \\ \hline
  \end{tabular}
\end{table}

Figure~\ref{Demomin} exemplifies the quasi-probability distribution function, $q_{s_1,s_2}(0,t_2)$, as a function of $\omega t_2$, where the left panel assumes $s_1=1,~s_2=-1$ and
the right panel assumes $s_1=-1, ~s_2=1$. 
For each curve in each panel, we fixed $\theta_0=0$ and $r$ as noted in the figure.
Each curve in Fig.~\ref{Demomin} adopted the parameters $x_0$ and $p_0$ noted in Table \ref{parameters}, which are chosen so that the quasi-probability distribution function multiplied $4$ achieves the maximum violation, $4q_{s_1,s_2}(0,t_2)=-0.113$.

With the use of Fig.~\ref{Demomin}, we explain how to understand the violation of the Leggett-Garg inequalities in an intuitive way. 
The violation of the Leggett-Garg inequalities appears when the position measurements give the opposite value against the expectation value.
Let us focus on the left panel of Fig.~\ref{Demomin}, which assumes $s_1=1$ and $s_2=-1$.
The initial values of $x_0$ and $p_0$ for the curves in the figure 
are roughly $-1.5<x_0<-0.5$ and $0.7<p_0<2$. 
This means that the initial expectation value of the coherent oscillation motion 
starts from the left $x_0<0$ with the right moving momentum $p_0>0$. $q_{1,-1}(0,t_2)$ computes the quasi-probability that the result of the position-measurement at $t_1=0$ gives $x>0$, which is opposite to the initial value of $x_0<0$,
and the result of the position-measurement at $t_2$ gives $x<0$. 
Figure \ref{HO} plots the expectation value of $\langle\psi_0|\hat x(t)|\psi_0\rangle/2|\xi|=\cos(\omega t-\Theta)$, 
where each curve assumes the same parameters adopted in the same type of curve in Fig.~\ref{Demomin}. 
From the left panel of Fig.~\ref{HO},
the expectation value of the position becomes positive after a short time. 
The quasi-probability distribution function $q_{1,-1}(0,t_2)$ takes the minimum negative values at the time $t_2$ with $\langle\psi_0|\hat x(t_2)|\psi_0\rangle >0$. This is opposite to the condition of  $q_{1,-1}(0,t_2)$, which computes the quasi-probability that the result of the position measurement at $t_2$ gives $x<0$. 
Thus the violation of the Leggett-Garg inequalities appears when the measurements give the opposite value against the expectation values. 

This is true for the right panel of Fig.~\ref{Demomin}, which plots $q_{-1,1}(0,t_2)$.
The initial values of $x_0$ and $p_0$ for the curves
in the figure are roughly $0.5<x_0<1.5$ and $0.7<p_0<2$.
This means that the initial expectation value of the coherent oscillation motion 
starts from the left $x_0>0$ with the right moving momentum $p_0>0$. 
$q_{-1,1}(0,t_2)$ computes the quasi-probability that the result of the position measurement at $t_1=0$ is $x<0$, which
 is opposite to the initial value of $x_0>0$, 
and the result of the position-measurement at $t_2$ gives $x>0$. 
As is shown in the right panel of Fig.~\ref{HO}, $\langle\psi_0|\hat x(t)|\psi_0\rangle/2|\xi|=\cos(\omega t-\Theta)$ becomes negative. The quasi-probability distribution function $q_{-1,1}(0,t_2)$ takes the minimum negative value at the time $t_2$ when the expectation value is negative $\langle\psi_0|\hat x(t_2)|\psi_0\rangle<0$. 
This is opposite to the condition of  $q_{-1,1}(0,t_2)$, which computes the quasi-probability that the result of the position measurement at $t_2$ gives $x>0$. 
Thus the violation of the Leggett-Garg inequalities occurs when the position measurements give the opposite values against the expectation values. 

Here we mention the reason why the curves in the panels of Fig.~\ref{Demomin} take the same values at $\omega t_2=\pi/2, ~\pi,~3\pi/2,~2\pi$. 
In subsection II~C, we showed that  
the quasi-probability distribution function 
$q_{s_1,s_2}(t_1,t_2)$
for a squeezed coherent state reduces to a quasi-probability distribution function for the coherent state by replacing the time $t_i$ and the initial parameter $\xi$ with $t_i+\beta(t_i)/\omega$ and $\xi'$, where $\beta(t)$ is defined by Eq.~(\ref{tprime}) and $\xi'$ is given by the relation in Eq.~(\ref{CS-SCStransform}). 
The quasi-probability distribution function for the squeezed coherent states adopted in each panel of Fig.~\ref{Demomin} reduces to the same 
quasi-probability distribution function for the coherent state by the transformation. We note that $\beta(t_2)=0$
when $\theta_0=0$ and $\omega t_2=\pi/2, ~\pi,~3\pi/2,~2\pi$.
This explains the reason why the quasi-probability distribution function
takes the same values at $\omega t_2=\pi/2, ~\pi,~3\pi/2,~2\pi$ in each panel of  Fig.~\ref{Demomin}. 

\subsection{Thermal squeezed coherent state }
In this subsection, we consider the thermal squeezed coherent state as the initial state, whose density matrix is given by
\begin{eqnarray}
    \hat{\rho}=\frac{1}{1+N_{th}}\sum^{\infty}_{m=0}\left(\frac{N_{th}}{1+N_{th}}\right)^mD(\alphax)^{\dagger}S(\zeta)^{\dagger}\Ket{m}\Bra{m}S(\zeta)D(\alphax),
\end{eqnarray}
where $N_{th}=\left[\exp({\hbar\omega}/{k_BT})-1\right]^{-1}$, and $T$ is the temperature. In this case, the explicit derivation of the quasi-probability is presented in Appendix B, which yields
\begin{eqnarray}
    q(s_1,s_2,t_1,t_2)&=&\frac{1}{1+N_{th}}\text{Re}\Biggr\{\frac{1}{4}\Bigl(1+s_1\mathrm{erf}(\frac{x_{\alphax(t_1)}}{\lambda(t_1)})+s_2\mathrm{erf}(\frac{x_{\alphax(t_2)}}{\lambda(t_2)})+s_1s_2\mathrm{erf}(\frac{x_{\alphax(t_1)}}{\lambda(t_1)})\mathrm{erf}(\frac{x_{\alphax(t_2)}}{\lambda(t_2)})\Bigr)\nonumber\\
    &\quad&+s_1s_2\Bigr\{\sum^{\infty}_{m=1}\left(\frac{N_{th}}{1+N_{th}}\right)^m e^{im\omega(t_2-t_1)+im(\beta(t_2)-\beta(t_1))}J_{m0}(-\frac{x_{\alphax(t_2)}}{\lambda(t_2)},\infty)J_{0m}(-\frac{x_{\alphax(t_1)}}{\lambda(t_1)},\infty)
    \nonumber\\
    &\quad&+\sum^{\infty}_{n=1}e^{-in\omega(t_2-t_1)-in(\beta(t_2)-\beta(t_1)}J_{0n}(-\frac{x_{\alphax(t_2)}}{\lambda(t_2)},\infty)J_{n0}(-\frac{x_{\alphax(t_1)}}{\lambda(t_1)},\infty)
    \nonumber\\
    &\quad&+\sum^{\infty}_{m=1}\sum^{\infty}_{n=1,m\neq n}\left(\frac{N_{th}}{1+N_{th}}\right)^mJ_{mn}(-\frac{x_{\alphax(t_2)}}{\lambda(t_2)},\infty)J_{nm}(-\frac{x_{\alphax(t_1)}}{\lambda(t_1)},\infty)\Bigr\}+ee(s_1,s_2,t_1,t_2)\Biggr\},
\end{eqnarray}
where the last term $ee(s_1,s_2,t_1,t_2)$ is defined depending on the values of $s_1$ and $s_2$, as follows,
\begin{eqnarray}
&&ee(s_1=1,s_2=1,t_1,t_2) =
\int^{\infty}_{-{x_{\alphax(t_2)}}/{\lambda(t_2)}}dx\sum^{\infty}_{n=1}\left(\frac{N_{th}}{1+N_{th}}\right)^n \psi^{\dagger}_n(x)\psi_n(x)\int^{\infty}_{-{x_{\alphax(t_1)}}/{\lambda(t_1)}}dy\psi^{\dagger}_n(y)\psi_n(y),   
\\
&&ee(s_1=1,s_2=-1,t_1,t_2) =
\int^{-{x_{\alphax(t_2)}}/{\lambda(t_2)}}_{-\infty}dx\sum^{\infty}_{n=1}\left(\frac{N_{th}}{1+N_{th}}\right)^n \psi^{\dagger}_n(x)\psi_n(x)\int^{\infty}_{-{x_{\alphax(t_1)}}/{\lambda(t_1)}}dy\psi^{\dagger}_n(y)\psi_n(y),
\\
&&ee(s_1=-1,s_2=1,t_1,t_2) =
\int^{\infty}_{-{x_{\alphax(t_2)}}/{\lambda(t_2)}}dx\sum^{\infty}_{n=1}\left(\frac{N_{th}}{1+N_{th}}\right)^n \psi^{\dagger}_n(x)\psi_n(x)\int^{-{x_{\alphax(t_1)}}/{\lambda(t_1)}}_{-\infty}dy\psi^{\dagger}_n(y)\psi_n(y),
\\
&&ee(s_1=-1,s_2=-1,t_1,t_2) =
\int^{-{x_{\alphax(t_2)}}/{\lambda(t_2)}}_{\infty}dx\sum^{\infty}_{n=1}\left(\frac{N_{th}}{1+N_{th}}\right)^n \psi^{\dagger}_n(x)\psi_n(x)\int^{-{x_{\alphax(t_1)}}/{\lambda(t_1)}}_{-\infty}dy\psi^{\dagger}_n(y)\psi_n(y).  
\end{eqnarray}

\begin{figure}[htbp]
    \begin{tabular}{cc}
          \begin{minipage}[t]{0.5\hsize}
        \centering
        \includegraphics[keepaspectratio, scale=0.48]{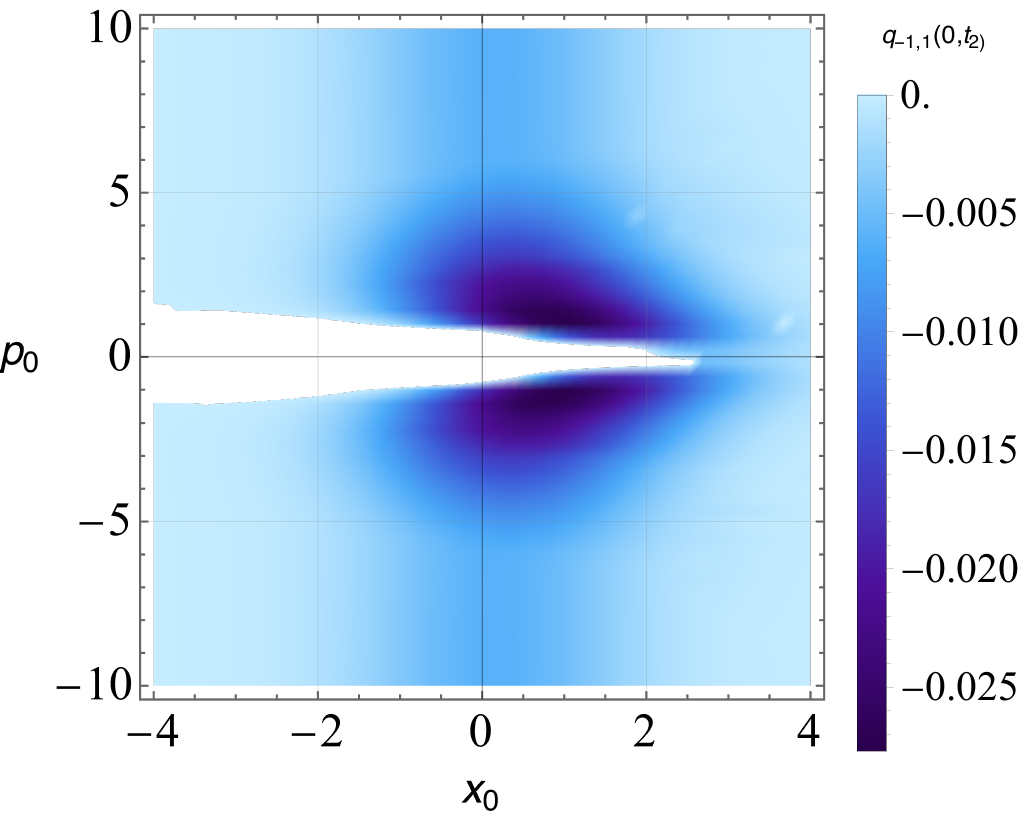}
        \subcaption{$k_BT/\hbar\omega=0$}
      \end{minipage} &
      \begin{minipage}[t]{0.5\hsize}
      \centering
    \includegraphics[keepaspectratio, scale=0.48]{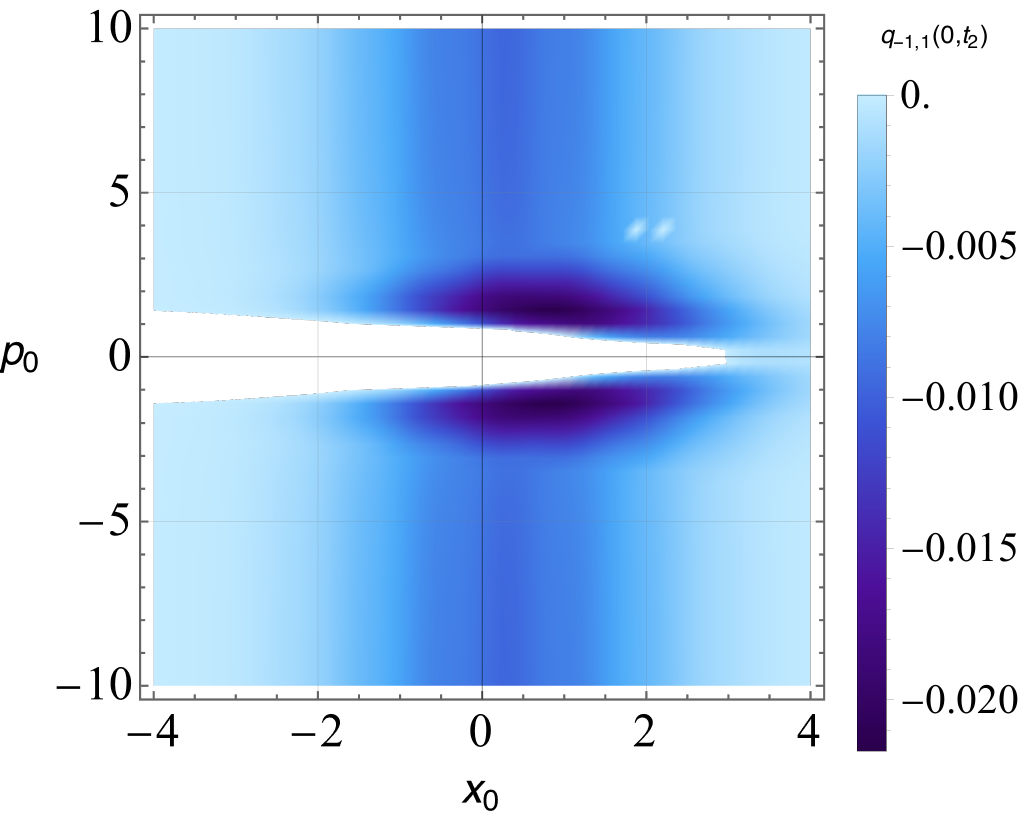}
        \subcaption{$k_BT/\hbar\omega=0.5$}
      \end{minipage} 
            \vspace{1cm}
      \\
     \begin{minipage}[t]{0.5\hsize}
        \centering
        \includegraphics[keepaspectratio, scale=0.48]{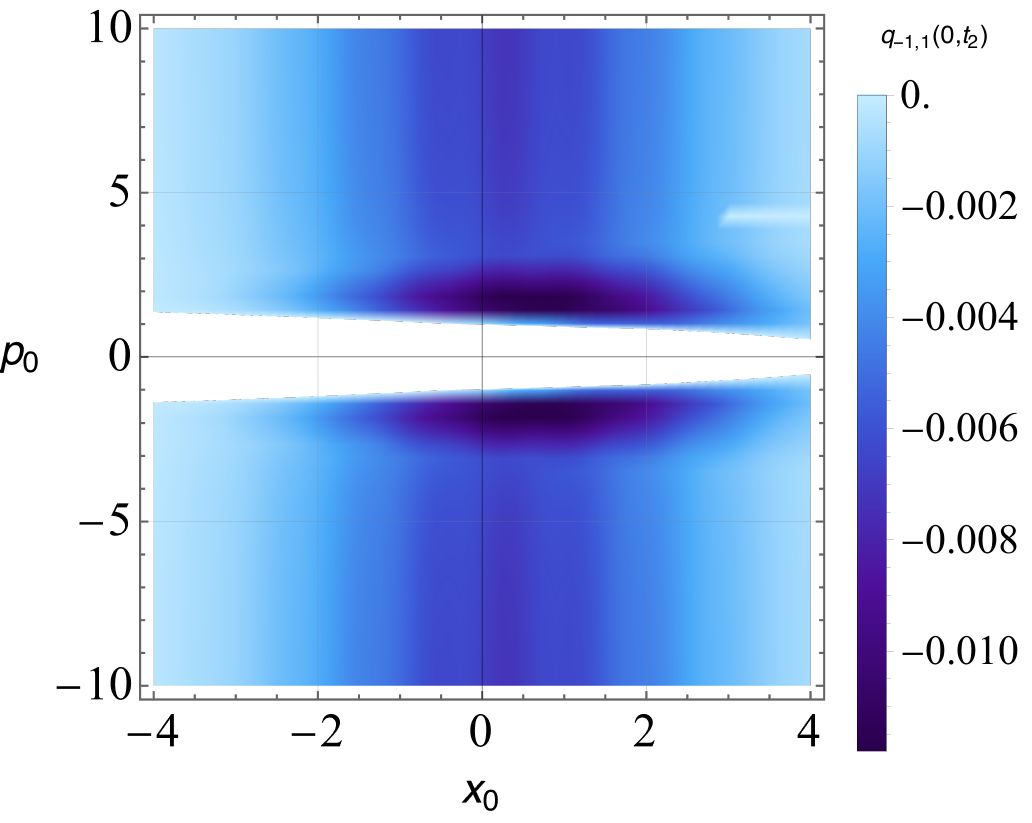}
        \subcaption{$k_BT/\hbar\omega=1.0$}
      \end{minipage} &
      \begin{minipage}[t]{0.5\hsize}
        \centering
        \includegraphics[keepaspectratio, scale=0.48]{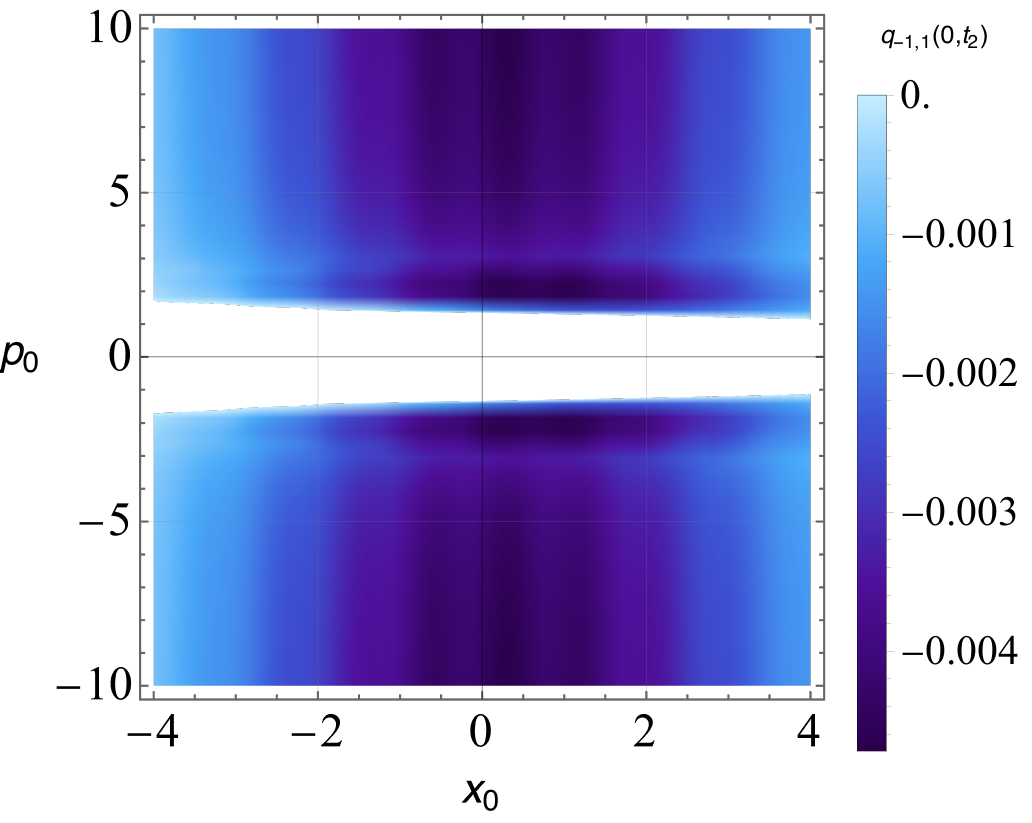}
        \subcaption{$k_BT/\hbar\omega=2$}
      \end{minipage} 
    \end{tabular}
     \caption{Contour plot of the minimum values of the two-time quasi-probability distribution function,
     $ q_{-1,1}(0,t_2)$, on the plane of $x_0$ and $p_0$. 
     Each panel assumes (a) $k_BT/\hbar \omega=0$,  (b) $k_BT/\hbar \omega=0.5$,
     (c) $k_BT/\hbar \omega=1.0$,  and (d) $k_BT/\hbar \omega=2$.  
     In each panel, we fixed $s_1=-1, s_2=1$ and the squeezing parameter
     $r=1/2$ and $\theta_0=0$.\label{pxpxT}}
  \end{figure}

The effect of the thermal state on the quasi-probability distribution function was discussed in Ref.~\cite{Halliwell23} focusing on the 
coherent state. Here we focus on the effect of squeezing. The panels of figure \ref{pxpxT} plot the minimum value of the
quasi-probability,  $q_{-1,1}(0,t_2)$, 
on the plane of $x_0$
and $p_0$, where we fixed the squeezing parameter $r=1/2$ and $\theta_0=0$.
 Each panel assumes
 (a) $k_BT/\hbar \omega=0$, 
  (b) $k_BT/\hbar \omega=0.5$, 
  (c) $k_BT/\hbar \omega=1$, 
  and (d) $k_BT/\hbar \omega=2$.
This figure demonstrates generally that the violation of 
the Leggett-Garg inequalities become weak as the temperature increases.

\subsection{Quasi-probability for the squeezed coherent state and coherent state}
In this section, we evaluate the effect of squeezing on the quasi-probability. 
The quasi-probability for a coherent state is written as \cite{Halliwell23}, 
\begin{eqnarray}
    q_{s_1,s_2}(t_1,t_2)&=&\frac{1}{4}\left\{1+s_1\mathrm{erf}\left(x_{\alphax(t_1)}\right)+s_2\mathrm{erf}\left(x_{\alphax(t_2)}\right)+s_1s_2\mathrm{erf}\left(x_{\alphax(t_1)}\right)\mathrm{erf}\left(x_{\alphax(t_2)}\right)\right\}\nonumber\\
&&\hspace{2pt}+s_1s_2\text{Re}\sum^{\infty}_{n=1}e^{-in\omega(t_2-t_1)}J_{0n}\left(x_{\alphax(t_1)},\infty\right)J_{n0}\left(x_{\alphax(t_2)},\infty\right).\label{quasi-probability of coherent state}
\end{eqnarray}
Comparing Eq.~(\ref{Hqsstt}) for the squeezed coherent state and 
Eq.~(\ref{quasi-probability of coherent state}) for the coherent state, the following 
relation between the two expressions holds. 
Namely, 
Eq.~(\ref{Hqsstt}) can be written 
using the quasi-probability for the 
coherent state Eq.~(\ref{quasi-probability of coherent state}) with replacing $\omega t_i$ by  $\omega t_i+\beta(t_i)$, where $i=1,~2$ and 
$\beta(t)$ is defined by Eq.~(\ref{tprime}). This can be read from 
\begin{eqnarray}
x_{\alphax(t_i+\beta(t_i)/\omega)}&=&\text{Re}[\sqrt{2}\xi e^{-i\omega t_i-i\beta(t_i)}]\nonumber\\
    &=&\text{Re}[(x_0+ip_0)(\cos({\omega t_i+\beta(t_i))}-i\sin{(\omega t_i+\beta(t_i))})]\nonumber\\
    &=&\frac{1}{\sqrt{(\cosh{r}+\cos{(\theta_0-2\omega t)}\sinh{r})^2+(\sin{(\theta_0-2\omega t)}\sinh{r})^2}}\nonumber\\
    &&\times \left\{x_0(\cos{\omega t}(\cosh{r}+\cos{(\theta_0-2\omega t_i)}\sinh{r})-\sin{\omega t_i}\sin{(\theta_0-2\omega t_i)}\sinh{r})\right.\nonumber
    \\
&&\hspace{20pt}\left.+p_0(\sin{\omega t_i}(\cosh{r}+\cos{(\theta_0-2\omega t_i)}\sinh{r})-\cos{\omega t_i}\sin{(\theta_0-2\omega t_i)}\sinh{r})\right\}\nonumber\\
    &=&\frac{x'_0\cos{\omega t_i}+p'_0\sin{\omega t_i}}{\lambda(t_i)},
\end{eqnarray}
where we defined
$\lambda(t_i)=\sqrt{\sinh (2r)\cos (2\omega t_i-\theta_0)+\cosh(2r)}$,~
$x'_0=x_0(\cosh{r}+\sinh{r}\cos{\theta_0})+p_0\sinh{r}\sin{\theta_0}$, 
~$p'_0=x_0\sinh{r}\sin{\theta_0}+p_0(\cosh{r}-\sinh{r}\cos{\theta_0})$.
Further, by defining $\xi'=(x_0'+ip_0')/\sqrt{2}$ we have
\begin{eqnarray}
x_{\alphax(t_i+\beta(t_i)/\omega)}&=&\text{Re}\biggl[\sqrt{2}{\xi' e^{-i\omega t_i}\over \lambda(t_i)}\biggr]=x_{\xi'(t_i)/\lambda(t_i)}.
\label{CS-SCStransform}
\end{eqnarray}
Thus, the quasi-probability distribution function for the squeezed coherent state 
reduces to the quasi-probability distribution function for the coherent state 
with the replacement $t_i\rightarrow t_i+\beta(t_i)/\omega$
and $\xi \rightarrow \xi'$.
This  explains that the maximum violation for the coherent state and the squeezed coherent state 
becomes equivalent when $t_2$ and $x_0$ and $p_0$ are taken free movable parameters. 
Such a relation was shown by using the identity
(\ref{SDrelation}) in Ref.~\cite{Halliwell23}. 
This same relation also holds for the thermal squeezed coherent state and the thermal coherent state. 

\begin{figure}[t]
    \begin{tabular}{cc}
          \begin{minipage}[t]{0.45\hsize}
        \centering
        \includegraphics[keepaspectratio, scale=0.45]{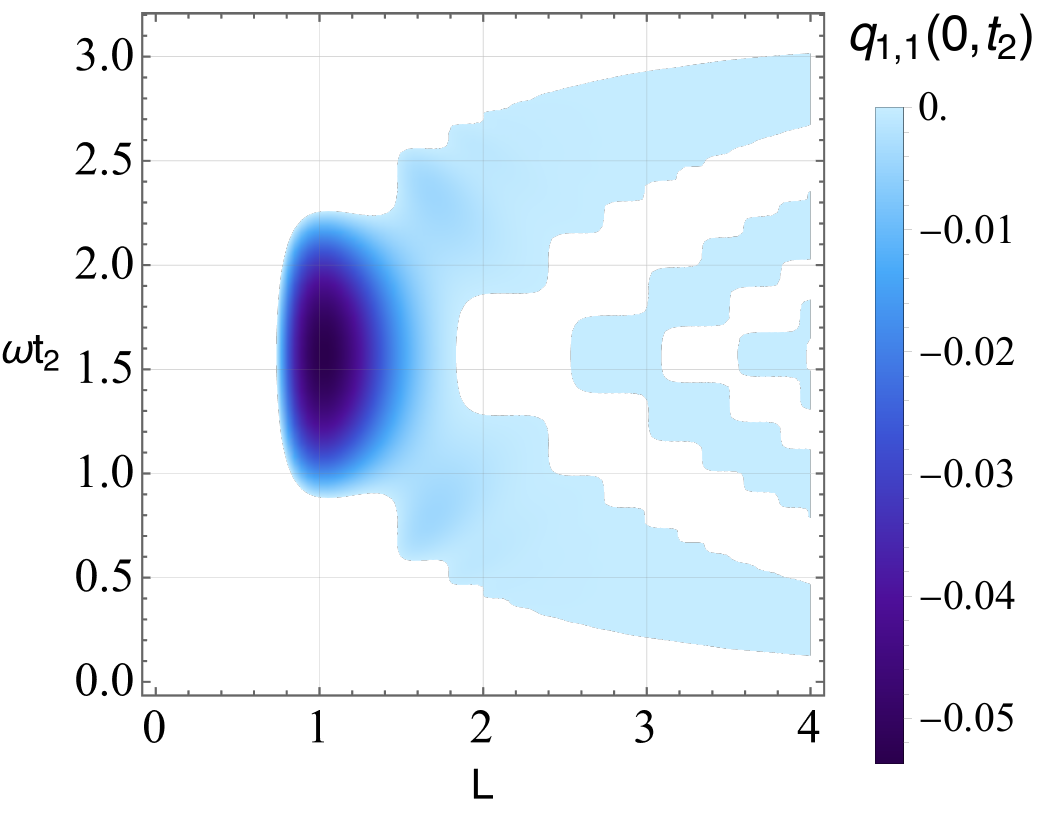}
      \end{minipage}
      &\hspace{0.5cm}
      \begin{minipage}[t]{0.45\hsize}
        \centering
        \includegraphics[keepaspectratio, scale=0.45]{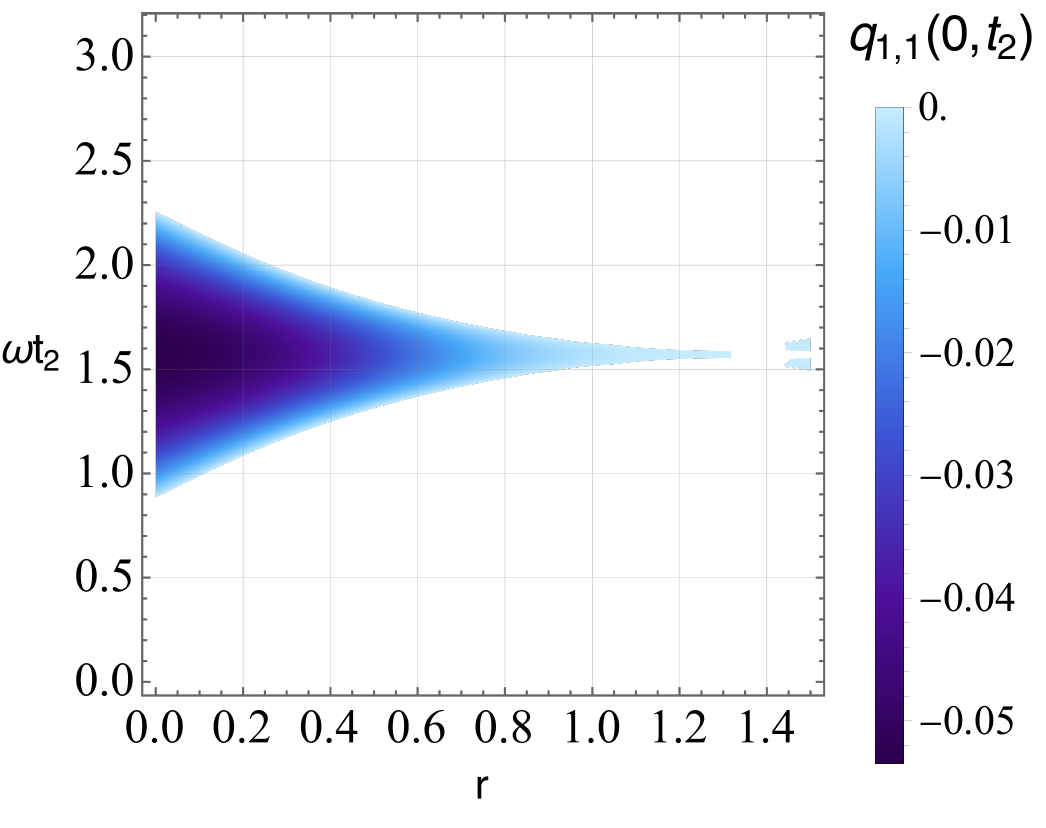}
      \end{minipage} 
        \end{tabular}
        \\

    \caption{Left panel is the contour of the quasi-probability $q_{1,1}(0,t_2)$ 
    of Eq.~(\ref{qssttL}) on the plane of $L$ and $\omega t_2$, 
    where we fixed $r=0$. 
    The right panel is the same but on the plane of $r$ and $\omega t_2$,  where we fixed $L=1$. 
    Each panel assumes  $\theta_0=0$, $s_1=1$, and $s_2=1$.}
    \label{pro_L_theta0}    
  \end{figure}

\begin{figure}
    \begin{tabular}{cc}
          \begin{minipage}[t]{0.5\hsize}
        \centering
        \includegraphics[keepaspectratio, scale=0.5]{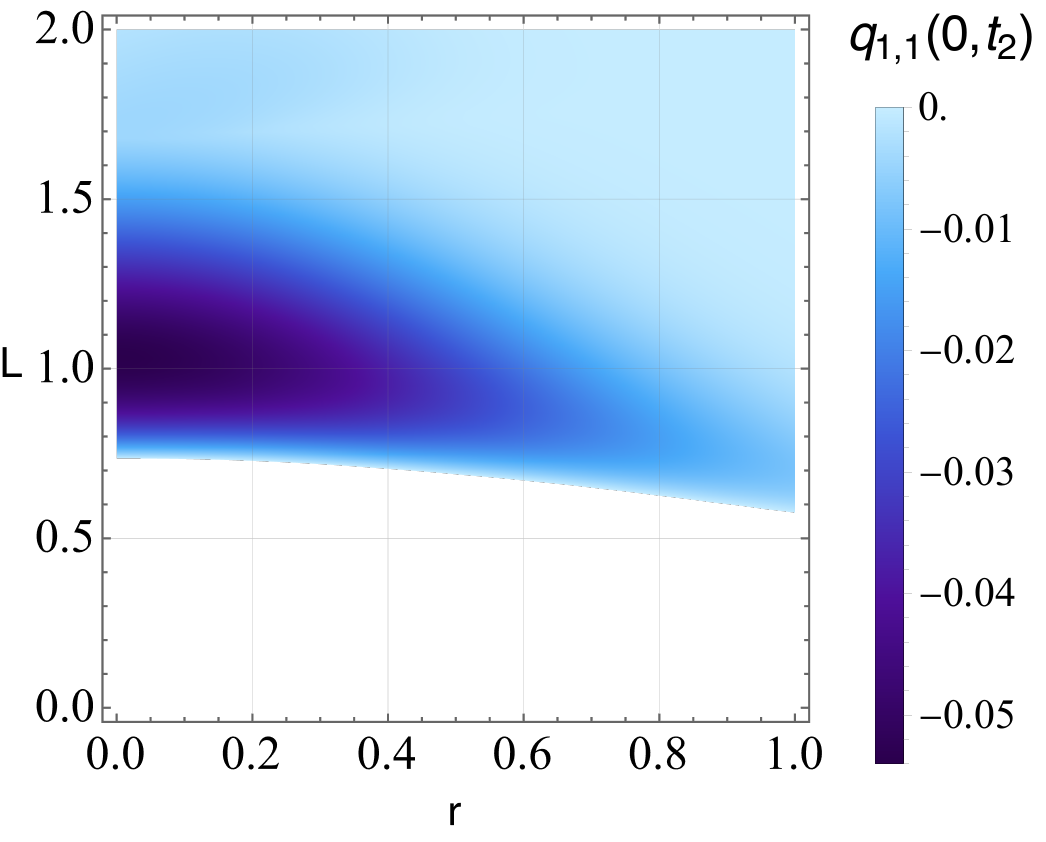}
      \end{minipage}
        \end{tabular}
        \\
    \caption{Contour of the minimum value of the quasi-probability $q_{1,1}(0,t_2)$  
    of Eq.~(\ref{qssttL}) on the plane of $r$ and $L$.
    Here $t_2$ takes different values at each point,
    but we fixed $\theta_0=0$.
    }
    \label{pro_L}
  \end{figure}

\begin{figure}
         \hspace{-5cm}
         \begin{minipage}[t]{0.4\hsize}
        \centering
     \includegraphics[keepaspectratio, scale=0.36]{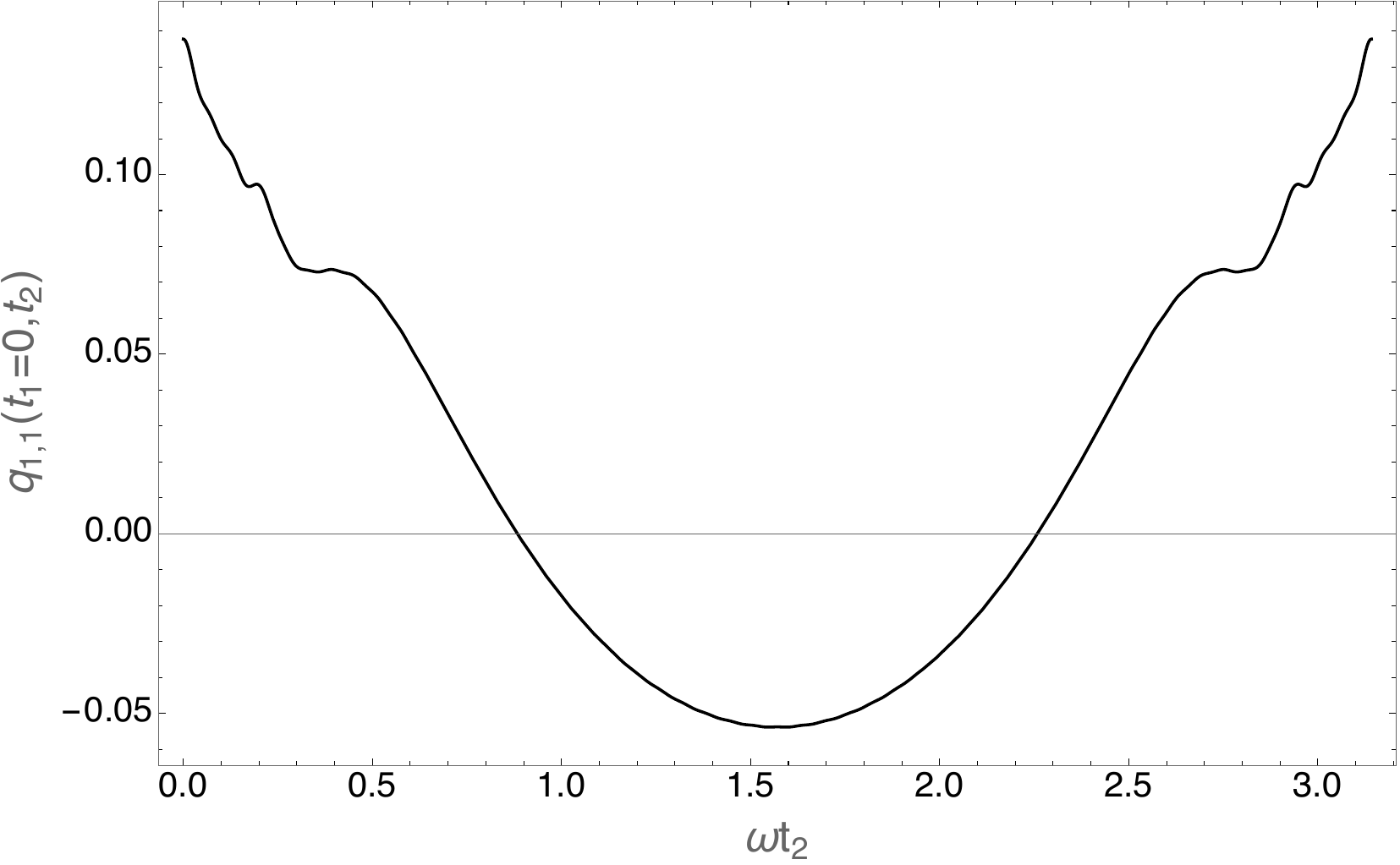}
      \end{minipage} 
    \caption{Quasi-probability $q_{1,1}(0,t_2)$  of Eq.~(\ref{qssttL})  as function of $\omega t_2$
    Here we fixed $r=0$, $\theta_0=0$, and $L=1.02$, which achieves the minimum value of the quasi-probability in Fig.~\ref{pro_L}. 
    }
    \label{proQP}
  \end{figure}
\section{A projection operator with a  larger LG violation}

In this section, we consider the violation of the two-time Leggett-Garg inequalities by adopting a different projection operator, for which 
we adopt the dichotomic variable $Q(\hat x)={\rm sgn}(\hat x-L)+{\rm sgn}(-\hat x-L)+1$, which leads to the projection operator given by
\begin{align}
    P_s
    &=\theta(s(\hat x-L))+\theta(-s(\hat x+L))+\frac{1}{2}(s-1).
\end{align}
The dichotomic variable defined by the above projection operator is
understood as follows. When the result of a measurement of the position of a harmonic oscillator is $|x|>L$, we assign $Q=1$. On the other hand,
when the result of a measurement of the position is $|x|\leq L$,
we assign $Q=-1$, where $L$ is a parameter of the operator. 
Therefore, the projection operator $P_s$ with $s=1$
gives the projection of the region $|x|>L$, while $P_s$ with $s=-1$
does the projection of the region $|x|\leq L$. 
In this section, we consider the squeezed state as the initial state,  $\rho_0=\ket{\zeta}\bra{\zeta}$, where $\ket{\zeta}$ is defined by $\ket{\zeta}=S(\zeta)\ket{0}$.
In this case, we evaluate the following formula as the quasi-probability
\begin{align}
    q_{s_1,s_2}(t_1,t_2)
    &=
    {\rm Re}\biggr[\bra{0}S^\dagger(\zeta) e^{iHt_2}\left\{ \theta(s_2(\hat x-L))+\theta(-s_2(\hat x+L))+\frac{1}{2}(s_2-1) \right\} e^{-iHt_2}\nonumber \\
    \label{quasip}
    &\times e^{iHt_1} \left\{ \theta(s_1(\hat x-L))+\theta(-s_1(\hat x+L))+\frac{1}{2}(s_1-1) \right\} e^{-iHt_1}S(\zeta)\ket{0}\biggr].
\end{align}
Using the method developed in the previous study in Ref.~\cite{Halliwell23}, we have
\begin{align}
      q_{s_1,s_2}(t_1,t_2)
      &=\frac{1}{4}\left\{ 1+s_1\left(1-2{\rm erf}\left(\frac{L}{\lambda(t_1)}\right)\right) \right\} \left\{ 1+s_2\left(1-2{\rm erf}\left(\frac{L}{\lambda(t_2)}\right)\right) \right\}\nonumber \\
      \nonumber\\
      &\hspace{-1cm}+s_1s_2{\rm Re}\Bigg[\sum^{\infty}_{n=1}e^{-in(\omega(t_2-t_1)+\beta(t_2)-\beta(t_1))} \Bigg\{ J_{0n}\left(\frac{L}{\lambda(t_1)},\infty\right)J_{n0}\left(\frac{L}{\lambda(t_2)},\infty\right)-J_{0n}\left(\frac{L}{\lambda(t_1)},\infty\right)J_{n0}\left(\frac{-L}{\lambda(t_2)},\infty\right)\nonumber \\
      &\hspace{60pt}-J_{0n}\left(\frac{-L}{\lambda(t_1)},\infty\right)J_{n0}\left(\frac{L}{\lambda(t_2)},\infty\right)+J_{0n}\left(\frac{-L}{\lambda(t_1)},\infty\right)J_{n0}\left(\frac{-L}{\lambda(t_2)},\infty\right)\Bigg\} \Bigg].
      \label{qssttL}
\end{align}

The panels of figure \ref{pro_L_theta0} plot the contour of the quasi-probability $q_{1,1}(0,t_2)$ of Eq.~(\ref{qssttL}) on the plane of $L$ and $\omega t_2$ with fixed $r=0$ (left panel), 
and on the plane of $r$ and $\omega t_2$ with fixed $L=1$ (right panel), where we fixed $\theta_0=0$ and  $s_1=s_2=1$. This is only the case that the Leggett-Garg inequality is clearly violated in the combinations of $s_1$ and $s_2$, although we found a very weak violation for $s_1=-1$ and $s_2=1$ with a nonzero value of $r$. 
The quasi-probability distribution function takes the negative values smaller than $-0.05$ for $L\sim 1$ and $r\simlt 0.5$ at  $\omega t_2 \sim \pi/2$. 
This number of the quasi-probability is smaller than those of the previous section, and the violation is boosted by the projection operator. 
The clear violation appears only when $s_1=s_2=1$, i.e., 
for the quasi-probability distribution function that the measurements at $t_1$ and $t_2$ give $|x|>L$. This is explained by the broadening feature of the wave function
coming from the superposition principle.

Figure \ref{pro_L} shows the minimum value of
the quasi-probability $q_{1,1}(0,t_2)$ on the plane of $L$ and $r$, where $t_2$ is the free parameter. 
Here we fixed $s_1=s_2=1$ and $\theta_0=0$. 
Figure~\ref{proQP} plots the quasi-probability $q_{1,1}(0,t_2)$ as function of $\omega t_2$, which achieves the minimum value of the quasi-probability in Fig.~\ref{pro_L}.
We note that the period of the quasi-probability of Eq.~(\ref{qssttL}) is $\pi/\omega$.
Figs.~\ref{pro_L_theta0} and \ref{pro_L} show that the relatively strong violation of the Leggett-Garg inequality appears for $0.8 \simlt L \simlt 1.2$ and $r\simlt 0.5$. 
Under the condition $\theta_0=0$,
the smallest value of the quasi-probability is $-0.0538$, 
which is $43\%$ of the L\"{u}ders bound. The smallest value appears for $r=0$ and $L=1.03$ at $\omega t_2 = 1.55$. 
Thus the choice of the dichotomic variable enables us to detect the violation of the Leggett-Garg inequality in the ground state and the squeezed state, 
which even boosts the violation of the amplitude.

\section{Summary and Conclusion}

In the present paper, we have investigated the violation of the two-time Leggett-Garg inequalities for testing the quantum nature of a harmonic oscillator in the various quantum states. 
We examined the two-time quasi-probability distribution function by constructing the dichotomic variable using the position operator of the harmonic oscillator in the squeezed coherent states and thermal squeezed coherent states, and demonstrated the violations of the Leggett-Garg inequalities by adopting a wide range of model parameters.
A useful choice of the dichotomic variable was developed to test the violations of the Leggett-Garg inequality for the ground state and squeezed states, which boosts the violation. 
We presented the two different formulas for the two-time quasi-probability distribution function.
One is the method of generalizing the previous work by Mawby and Halliwell \cite{Halliwell23}, while the other is the one newly developed using the integral formula of the Heaviside function. We have demonstrated that both formulas produce the same result for the squeezed coherent state. 
These two formulas have merit and demerit. 
For example, using the method generalized in the previous work \cite{Halliwell23}, it was easy to compute the thermal squeezed coherent state as demonstrated in Sec.~3.
On the other hand, the former method 
in Sec.~2 using the integral formula of the Heaviside function can be generalized to a formulation for a field theory, which will be reported in~\cite{Tani,Hirotani}.

The explicit computations of the quasi-probability distribution function have demonstrated that the minimum value is the same for the coherent states and the squeezed coherent states. Therefore, as predicted in the previous paper in Ref.~\cite{Halliwell23}, squeezing coherent states does not increase the violation of the Leggett-Garg inequalities in our model.
We have also demonstrated how the thermality weakens the violation of the Leggett-Garg inequalities. 
For the thermal squeezed coherent state, the minimum value of the quasi-probability distribution function becomes larger, and then it becomes positive as the temperature increases. This is understood as the state approaches classical states as the temperature increases. 
We have explicitly demonstrated that the quasi-probability distribution function of  the (thermal) squeezed coherent state reduces to the quasi-probability distribution function for the (thermal) coherent state by replacing
the parameters $t_1$ and $t_2$ and $\xi$, which explains that the common features in the (thermal) squeezed coherent state and the (thermal) coherent states.

We have also demonstrated that the violation of the Leggett-Garg inequalities occurs even for the ground state and the squeezed state by choosing the dichotomic variable in non-trivial ways. For example, 
$Q={\rm sgn}(x-\bar x(t))$
predicts the violation of the Leggett-Garg inequalities for the ground state by choosing a time-dependent function $\bar x(t)$. 
Further, $Q={\rm sgn}(x-L)+{\rm sgn}(-x-L)+1$
predicts the larger violation of the Leggett-Garg inequalities for the ground state and the squeezed state. Namely, the smaller negative values of the two-time quasi-probability distribution function are obtained by adopting $Q={\rm sgn}(x-L)+{\rm sgn}(-x-L)+1$ for the ground state and the squeezed state than the simple case of $Q={\rm sgn}(x-\bar x(t))$ for the squeezed coherent state. Measurement operators for much larger violations of the Leggett-Garg inequalities will be discussed in \cite{Hirotani}.

It is known that the two-time Leggett-Garg inequalities and the three-time Leggett-Garg 
inequalities are necessary and sufficient to prove macrorealism. 
Therefore, it might be useful to investigate the three-time 
Leggett-Garg inequalities.  
Application of the Leggett-Garg inequalities to realistic 
optomechanical experiments \cite{Matsumoto,MY} should be investigated in the future.
To this end, we further need to extend the formulation for the system taking the impacts of noises of environments, feedback control, and quantum filtering process into account. The method to realize the projection operators assumed in the present paper is left as a future investigation.

	\acknowledgements
We thank 
 Yuki Osawa, Youka Kaku, Yasusada Nambu, Masahiro Hotta, Akira Matsumura,  Nobuyuki Matsumoto for the discussions on this topic in the present work.
	K.Y. was supported by JSPS KAKENHI (Grant No.~JP22H05263. and No.~JP23H01175).
	D. M. was supported by JSPS KAKENHI (Grant No.~JP22J21267). 
\appendix
\section{Two-time quasi-probability for the squeezed coherent state(Mawby and Halliwell's method)}
In this Appendix, we derive the expression for the two-time quasi-probability (\ref{Hqsstt}).
The derivation in this appendix is almost identical to those in Ref.\cite{Halliwell23} for the coherent states, but the difference 
comes from the squeezing operator. We start with the expression
\begin{eqnarray}
    q_{s_1,s_2}(t_1,t_2)=\text{Re}[\Bra{0}S^{\dagger}(\zeta)D^{\dagger}(\alphax)e^{i\hat{H}t_2}\theta(s_2\hat{x})e^{-i\hat{H}(t_2-t_1)}
 \theta(s_1\hat{x})e^{-i\hat{H}t_1}D(\alphax)S(\zeta)\Ket{0}],
 \label{Aqss}
\end{eqnarray}   
adopting the initial state of the harmonic oscillator in the squeezed coherent state, $\rho_0=D(\alphax)S(\zeta)|0\rangle \langle0|S^\dagger(\zeta)D^\dagger(\alphax)$. 
Using the formula, 
 \begin{eqnarray}
 e^{-i\hat{H}t}D(\alphax)S(\zeta)=D(\alphax(t))S(\zeta(t))e^{-i\hat{H}t}
 \end{eqnarray}   
 with defined $\alphax(t)=\alphax e^{-i\omega t}$ and $\zeta(t)=\zeta e^{-2i\omega t}$, the right hand side of Eq.~(\ref{Aqss}) is written as
\begin{eqnarray}
    q_{s_1,s_2}(t_1,t_2)=\text{Re}\Bra{0}e^{i\hat{H}t_2}S^{\dagger}(\zeta(t_2))D^{\dagger}(\alphax(t_2))\theta(s_2\hat{x})e^{-i\hat{H}(t_2-t_1)}\theta(s_1\hat{x})D(\alphax(t_1))S(\zeta(t_1))e^{-i\hat{H}t_1}\Ket{0}.
\end{eqnarray}
Further, since we can write
\begin{eqnarray}
&&\theta(s\hat{x})D(\alphax(t))=D(\alphax(t))\theta(s(\hat x+x_{\alphax(t)})),
\\
&&S(\zeta(t_2))^{\dagger}e^{-i\hat{H}(t_2-t_1)}S(\zeta(t_1))=S(\zeta(t_2))^{\dagger}S(\zeta(t_2))e^{-i\hat{H}(t_2-t_1)}=e^{-i\hat{H}(t_2-t_1)}
\\
&& 
D(\alphax(t_2))^{\dagger}e^{-i\hat{H}(t_2-t_1)}D(\alphax(t_1))=D(\alphax(t_2))^{\dagger}D(\alphax(t_2))e^{-i\hat{H}(t_2-t_1)}=e^{-i\hat{H}(t_2-t_1)},  
\end{eqnarray}
where $x_{\alphax(t)}=\sqrt{2}\text{Re}[\alphax(t)]$, 
we have
\begin{eqnarray}
    q_{s_1,s_2}(t_1,t_2)=\text{Re}[e^{{i\omega(t_2-t_1)}/{2}}\Bra{0}S^{\dagger}(\zeta(t_2))\theta(s_2(\hat{x}+x_{\alphax(t_2)}))e^{-i\hat{H}(t_2-t_1)}\theta(s_1(\hat{x}+x_{\alphax(t_1)}))S(\zeta(t_1))\Ket{0}].
\end{eqnarray}

Using the properties of the unitary operator $S(\zeta)$ and the Bogoliubov transformation, we have
\begin{eqnarray}
    \theta(s(\hat{x}+x_{\alphax(t)}))S(\zeta(t))
    &=&S(\zeta(t))S^{\dagger}(\zeta(t))\theta(s(\hat{x}+x_{\alphax(t)}))S(\zeta(t))\nonumber \\
    &=&S(\zeta(t))S^{\dagger}(\zeta(t))\theta(s(\frac{\hat{a}+\hat{a}^{\dagger}}{\sqrt{2m\omega}}+x_{\alphax(t)}))S(\zeta(t))\nonumber \\
    &=&S(\zeta(t))\theta\Bigl(s(\frac{\hat{a}\cosh{r}+\hat{a}^{\dagger}e^{i\theta}\sinh{r}+\hat{a}^{\dagger}\cosh{r}+\hat{a}e^{-i\theta}\sinh{r}}{\sqrt{2m\omega}}+x_{\alphax(t)})\Bigr)\nonumber \\
    &=&S(\zeta)\theta\Bigl(s(A(t)\hat{x}+B(t)\frac{\hat{p}}{m\omega}+x_{\alphax(t)})\Bigr),
    \label{definitionbeta}
\end{eqnarray}
where we defined 
\begin{eqnarray}
A(t)&=&\cosh{r}+\cos{(\theta_0-2\omega t)}\sinh{r},~~~~
    B(t)=\sin{(\theta_0-2\omega t)}\sinh{r}, 
\end{eqnarray}
and $\zeta=re^{i\theta_0}$.
Next, consider a polar coordinate transformation of the linear combination of the position and momentum operators in phase space.
We defined $\lambda(t)=\sqrt{A(t)^2+B(t)^2}=\sqrt{\sinh (2r)\cos (2\omega t-\theta_0)+\cosh(2r)}$, $A(t)$ and $B(t)$ can be rewritten as
\begin{eqnarray}
    A(t)&=&\lambda(t)\cos{\beta(t)},~~~~
    B(t)=\lambda(t)\sin{\beta(t)},\\
    \beta(t)&=&\arctan\left(\frac{B(t)}{A(t)}\right)
        =\arctan\left[{\frac{\sin{(\theta_0-2\omega t)}\sinh{r}}{\cosh{r}+\cos{(\theta_0-2\omega t)\sinh{r}}}}\right].\label{tprime}
\end{eqnarray}
Then, using 
\begin{eqnarray}
    A(t)\hat{x}+B(t)\frac{\hat{p}}{m\omega}&=&\lambda(t)(\hat{x}\cos{\beta(t)}+\frac{\hat{p}}{m\omega}\sin{\beta(t)})
=\lambda(t)\frac{\hat{a}e^{-i\beta(t)}+\hat{a}^{\dagger}e^{i\beta(t)}}{\sqrt{2m\omega}}
 =   \lambda(t)e^{i\hat{H}\beta(t)/\omega}\hat{x}e^{-i\hat{H}\beta(t)/\omega}=\lambda(t)\hat x(\beta(t)),
 \nonumber\\
\end{eqnarray}
and $\theta(s(\lambda(t)e^{i\hat{H}\beta(t)/\omega}\hat{x}e^{-i\hat{H}\beta(t)/\omega}+x_{\alphax(t)}))=e^{i\hat{H}\beta(t)/\omega}\theta(s(\lambda(t)\hat{x}+x_{\alphax(t)}))e^{-i\hat{H}\beta(t)/\omega}$, we have the quasi-probability 
\begin{eqnarray}
&&q_{s_1,s_2}(t_1,t_2)\nonumber\\
&\quad&=\text{Re}
\biggl[e^{i\omega(t_2-t_1)/2}\Bra{0}\theta(s_2(\lambda(t_2)e^{i\hat{H}\frac{\beta(t_2)}{\omega}}\hat{x}e^{-i\hat{H}\frac{\beta(t_2)}{\omega}}+x_{\alphax(t_2)}))e^{-i\hat{H}(t_2-t_1)}\theta(s_1(\lambda(t_1)e^{i\hat{H}\frac{\beta(t_1)}{\omega}}\hat{x}e^{-i\hat{H}\frac{\beta(t_1)}{\omega}}+x_{\alphax(t_1)}))\Ket{0}\biggr]\nonumber \\
&\quad&=\text{Re}\biggl[e^{i\omega\frac{(t_2-t_1)}{2}}e^{i\frac{\beta(t_2)-\beta(t_1)}{2}}\sum^{\infty}_{n=0}\Bra{0}\theta(s_2(\lambda(t_2)\hat{x}+x_{\alphax(t_2)}))e^{-i\hat{H}\frac{\beta(t_2)}{\omega}}e^{-i\hat{H}(t_2-t_1)}e^{i\hat{H}\frac{\beta(t_1)}{\omega}}\Ket{n}\Bra{n}\theta(s_1(\lambda(t_1)\hat{x}+x_{\alphax(t_1)}))\Ket{0}\biggr]\nonumber\\
&\quad&=\text{Re}\biggl[\sum^{\infty}_{n=0}e^{-in\omega(t_2-t_1)}e^{-in(\beta(t_2)-\beta(t_1))}\Bra{0}\theta\left(s_2(\hat{x}+\frac{x_{\alphax(t_2)}}{\lambda(t_2)})\right)\Ket{n}\Bra{n}\theta\left(s_1(\hat{x}+\frac{x_{\alphax(t_1)}}{\lambda(t_1)})\right)\Ket{0}\biggr],
\end{eqnarray}
which leads to Eq.~(\ref{Hqsstt}).

\section{Two-time quasi-probability for the thermal squeezed coherent state}
Here we derive the expression for the quasi-probability distribution 
when the initial state is the thermal squeezed coherent state, 
whose density matrix is written as
\begin{eqnarray}
    \hat{\rho}=\frac{1}{1+N_{th}}\sum^{\infty}_{m=0}\left(\frac{N_{th}}{1+N_{th}}\right)^mD(\alphax)S(\zeta)\Ket{m}\Bra{m}S(\zeta)^{\dagger}D(\alphax)^{\dagger},
\end{eqnarray}
where $N_{th}=\left[\exp({\hbar\omega}/{k_BT})-1\right]^{-1}$, and 
$T$ is the temperature. In this case, the two-time quasi-probability is given by
\begin{eqnarray}
  q_{s_1,s_2}(t_1,t_2)=\frac{1}{1+N_{th}}\sum^{\infty}_{m=0}\left(\frac{N_{th}}{1+N_{th}}\right)^m\text{Re}\biggl[\Bra{m}S^{\dagger}(\zeta)D^{\dagger}(\alphax)e^{i\hat{H}t_2}\theta(s_2\hat{x})e^{-i\hat{H}(t_2-t_1)}\theta(s_1\hat{x})e^{-i\hat{H}t_1}D(\alphax)S(\zeta)\Ket{m}\biggr].
  \nonumber\\
\end{eqnarray}
By using the result in the Appendix A, we have

\begin{eqnarray}
    &&q_{s_1,s_2}(t_1,t_2)\nonumber \\
    &=&\frac{1}{1+N_{th}}\sum^{\infty}_{m,n=0}\left(\frac{N_{th}}{1+N_{th}}\right)^m\nonumber\\
   &\times&\text{Re}\biggl[\Bra{m}e^{i\hat{H}t_2+i\hat{H}\frac{\beta(t_2)}{\omega}}\theta\left(s_2(\hat{x}+\frac{x_{\alphax(t_2)}}{\lambda(t_2)})\right)e^{-i\hat{H}\frac{\beta(t_2)-\beta(t_1)}{\omega}-i\hat{H}(t_2-t_1)}\Ket{n}\Bra{n}\theta\left(s_1(\hat{x}+\frac{x_{\alphax(t_1)}}{\lambda(t_1)})\right)e^{-i\hat{H}\frac{\beta(t_1)}{\omega}-i\hat{H}t_1}\Ket{m}\biggr]\nonumber \\
    &=&\frac{1}{1+N_{th}}\sum^{\infty}_{m,n=0}\left(\frac{N_{th}}{1+N_{th}}\right)^m\nonumber \\
    &\times&\text{Re}\biggl[e^{i(m-n)(\omega(t_2-t_1)+\beta(t_2)-\beta(t_1))}\Bra{m}\theta\left(s_2(\hat{x}+\frac{x_{\alphax(t_2)}}{\lambda(t_2)})\right)\Ket{n}\Bra{n}\theta\left(s_1(\hat{x}+\frac{x_{\alphax(t_1)}}{\lambda(t_1)})\right)\Ket{m}\biggr]\nonumber\\
    &=&\frac{1}{1+N_{th}}\sum^{\infty}_{m=0}\left(\frac{N_{th}}{1+N_{th}}\right)^m\text{Re}\biggl[e^{im(\omega(t_2-t_1)+\beta(t_2)-\beta(t_1))}\Bra{m}\theta\left(s_2(\hat{x}+\frac{x_{\alphax(t_2)}}{\lambda(t_2)})\right)\Ket{0}\Bra{0}\theta\left(s_1(\hat{x}+\frac{x_{\alphax(t_1)}}{\lambda(t_1)})\right)\Ket{m}\nonumber\\
    &\quad&+\sum^{\infty}_{n=1}e^{i(m-n)(\omega(t_2-t_1)+\beta(t_2)-\beta(t_1))}\Bra{m}\theta\left(s_2(\hat{x}+\frac{x_{\alphax(t_2)}}{\lambda(t_2)})\right)\Ket{n}\Bra{n}\theta\left(s_1(\hat{x}+\frac{x_{\alphax(t_1)}}{\lambda(t_1)})\right)\Ket{m}\biggr].
\end{eqnarray}

By separating the term $m=0$ from the other terms of $m\neq0$ in the sum of $m$, we have
\begin{eqnarray}
      &&q_{s_1,s_2}(t_1,t_2)\nonumber\\
      &=&\frac{1}{1+N_{th}}\text{Re}\Biggl[\Bra{0}\theta\left(s_2(\hat{x}+\frac{x_{\alphax(t_2)}}{\lambda(t_2)})\right)\Ket{0}\Bra{0}\theta\left(s_1(\hat{x}+\frac{x_{\alphax(t_1)}}{\lambda(t_1)})\right)\Ket{0}\nonumber\\
    &\quad&+\sum^{\infty}_{m=1}\left(\frac{N_{th}}{1+N_{th}}\right)^me^{im(\omega(t_2-t_1)+\beta(t_2)-\beta(t_1))}\Bra{m}\theta\left(s_2(\hat{x}+\frac{x_{\alphax(t_2)}}{\lambda(t_2)})\right)\Ket{0}\Bra{0}\theta\left(s_1(\hat{x}+\frac{x_{\alphax(t_1)}}{\lambda(t_1)})\right)\Ket{m}\nonumber\\
    &\quad&+\sum^{\infty}_{n=1}e^{-in(\omega(t_2-t_1)+\beta(t_2)-\beta(t_1))}\Bra{0}\theta\left(s_2(\hat{x}+\frac{x_{\alphax(t_2)}}{\lambda(t_2)})\right)\Ket{n}\Bra{n}\theta\left(s_1(\hat{x}+\frac{x_{\alphax(t_1)}}{\lambda(t_1)})\right)\Ket{0}\nonumber\\
    &\quad&+\sum^{\infty}_{m=1}\sum^{\infty}_{n=1}\left(\frac{N_{th}}{1+N_{th}}\right)^me^{i(m-n)(\omega(t_2-t_1))+\beta(t_2)-\beta(t_1))}\Bra{m}\theta\left(s_2(\hat{x}+\frac{x_{\alphax(t_2)}}{\lambda(t_2)})\right)\Ket{n}\Bra{n}\theta\left(s_1(\hat{x}+\frac{x_{\alphax(t_1)}}{\lambda(t_1)})\right)\Ket{m}\Biggr].\nonumber\\
    &&\hspace{300pt}
    \label{TSCS}
\end{eqnarray}

\end{document}